\documentstyle[aps,epsf]{revtex} 
\draft

\begin{document}
\voffset 0.5 true in

\title{Slowly Rotating General Relativistic Superfluid Neutron Stars}
\author{N. Andersson}
\address{Department of Mathematics, University of Southampton\\ 
Southampton, UK}
\author{G. L. Comer}
\address{Department of Physics, Saint Louis University\\
P.O. Box 56907, St. Louis, MO 63156-0907, USA}

\date{\today}

\def\beq{\begin{equation}}
\def\eeq{\end{equation}}
\def\n{n}
\def\p{p}
\def\d{\delta}
\def\a00{{{\cal A}_0^0}}
\def\b00{{{\cal B}_0^0}}
\def\c00{{{\cal C}_0^0}}
\def\D00{{{\cal D}_0^0}}
\def\A{{\cal A}}
\def\B{{\cal B}}
\def\C{{\cal C}}

\maketitle

\begin{abstract}
We present a general formalism to treat  slowly rotating general 
relativistic superfluid neutron stars.  As a first approximation, 
their matter content can be described in terms of a two-fluid 
model, where one fluid is the neutron superfluid, which is believed 
to exist in the core and inner crust of mature neutron stars, and the 
other fluid represents a conglomerate of all other constituents (crust 
nuclei, protons, electrons, etc.).  We obtain a system of equations,  
good to second-order in the rotational velocities, that determines the 
metric and the matter variables, irrespective of the equation of state 
for the two fluids.  In particular, allowance is made for the 
so-called entrainment effect, whereby the momentum of one constituent 
(e.g. the neutrons) carries along part of the mass of the other 
constituent.  As an illustration of the developed framework, 
we consider a simplified equation of state for which the two fluids are 
described by different polytropes.  We determine numerically the effects 
of the two fluids on the rotational frame-dragging, the induced changes 
in the neutron and proton densities and 
the inertial mass,  as well as the change in shape of the star.  
We further discuss issues regarding conservation of the 
two baryon numbers, the mass-shedding (Kepler) limit and chemical
equilibrium.
\end{abstract}

\section{Introduction}

For very practical reasons, past modeling of general relativistic 
neutron stars has largely relied on a one-constituent perfect fluid 
approximation to describe the stress-energy of matter. This 
would be appropriate if neutron stars truly consisted solely
of neutrons in a fluid state.   However, this is a drastic 
oversimplification. In reality a neutron star is composed of many 
different constituents and is best described as a ``layer-cake''. 
A few seconds after it is 
born in a supernova a neutron star cools below
$10^{10}$~K. Then the outer layers (with densities below,
say, $1.5\times 10^{14}$~g/cm$^3$) will ``freeze'' into a solid crust.
After a few months of further cooling, the star will reach the 
critical temperature at which the bulk of the neutrons at densities
above neutron drip ($\sim 10^{11}$~g/cm$^3$) become superfluid. 
At roughly the same time the protons (expected to make up a few percent
of the core material) may become superconducting. 
Thus, even in the simplest reasonable 
model a neutron star contains a thin fluid ocean, 
the  
roughly 1~km deep solid crust mentioned above, and a core containing 
superfluid neutrons, (potentially superconducting) protons and electrons.
The lattice of nuclei and 
electrons in the inner crust is permeated by superfluid neutrons.
Yet more complex models would allow for the presence of exotic
particles in the core (eg. hyperons),  
high density phase-transitions
to  kaon and/or pion condensates and/or regions with deconfined
quarks~\cite{glendenning}. 

Despite these many complexities the perfect fluid approximation 
is reasonable when all 
the constituents flow together (see, for instance, Comer and Langlois 
\cite{GCL2}). However,  it can never 
accomodate situations where
the superfluid neutrons flow ``independently'' of the protons.
Since this is, in fact, expected to be the case in a real 
neutron star the standard approach has serious deficiencies.
During the last twenty years there have  been continued
attempts to piece together a coherent
picture of neutron star superfluidity, using 
a variety of theoretical arguments and imposing observational
constraints based on glitch and cooling rate data. Of particular interest
for our present discussion  are issues regarding the coupling between 
the superfluid and the normal fluid constituents, the main 
question being whether the superfluid neutrons will co-rotate
with the protons or not. The answer to this seemingly innocent 
question is very complicated, but the  current consensus is that even 
though the two fluids are strongly coupled they can sustain a state 
of differential rotation over significant timescales.
  
Alpar et al \cite{ALS} have shown that the charged components in the 
core fluid (the protons and the electrons) will be strongly coupled to 
the crust nuclei via electromagnetic interactions. This suggests that 
a superfluid star can be described in terms of two fluids: 
one that represents the superfluid neutrons and one that 
describes all charged components (protons, electrons, 
crust nuclei etc.). In the following we will refer to these two
fluids as the ``neutrons'' and the ``protons'', respectively. 
Furthermore, in a rotating star the normal fluid constituent will 
settle into an essentially non-dissipative configuration
on the viscous timescale. This implies that the ``protons'' must 
rotate uniformly (with  proton angular velocity 
$\Omega_\p=$~constant).
This is not a priori the case for the neutrons since a (pure) 
superfluid is non-dissipative. In other words, one can conceivably
have a stationary configuration with differentially rotating 
neutrons (with neutron angular velocity 
$\Omega_n\neq$~constant). This would certainly be true if the
two fluids were completely uncoupled but, in reality, this is 
likely not the case. In order to begin to understand what 
really happens we need to 
discuss a non-dissipative mechanism known as the entrainment effect 
and how it leads to a dissipative mechanism called mutual friction.

In a mixture of the two superfluids Helium three and 
Helium four, it is known that the momentum of one of the constituents 
carries along (or entrains) some of the mass of the 
other constituent 
\cite{AB,TT}.  The analog in neutron stars is an entrainment of some of 
the protons, say, by the neutrons. A superfluid is by its very nature
locally irrotational \cite{P}, so when neutron stars rotate they 
do so via the formation of a dense array of quantized vortices.
Because of entrainment 
the flow of neutrons around these vortices will 
induce a flow also in a fraction of the protons.  This leads to magnetic 
fields being formed around the vortices.  Due to the 
electromagnetic attraction of the electrons to the protons (the 
timescale of which is very short) there will then be a dissipative 
scattering of the electrons. In the context of oscillating
superfluid stars this dissipative mechanism is known as  
mutual friction. 

Much of the discussion of neutron star superfluids concerns attempts
to theoretically model the observed glitches. 
The favoured model has long been based on the idea that 
the glitches correspond to a transfer of angular momentum from 
the superfluid that permeates the crust to the lattice. This
picture is attractive since 
$$
{\Delta \dot{\Omega} \over \dot{\Omega} } \sim {\Delta I \over I} \sim
10^{-2}-10^{-3}
$$ 
where $\Delta \dot{\Omega}/\dot{\Omega}$ is the observed relative
change in spindown rate
and $\Delta I/I$ is the fraction of the total moment of inertia 
contained in the crustal superfluid. Furthermore, it has been argued
that the time-scale on which the core superfluid couples to 
the charged component 
(in our case, the time on which neutron rotation rate becomes 
equal to that of the protons) is short (on the order of 1-10~s).  
For example, Alpar and Sauls~\cite{AS1} argue that, 
if $P$ is the period of rotation 
of the bulk of the matter of the neutron star in units of seconds, 
then on timescales of $4 \times 
10^2-10^4~P$ any difference of rotation between the neutrons and protons 
in the core will be damped out because of 
the entrainment effect and mutual friction. The consequence then is that 
only the inner crust superfluid would be able to rotate at 
a different rate from the rest of the star on a timescale
beyond at most a few minutes.

However, there are many uncertainties in this game, 
e.g. the strength of the vortex pinning and the exact parameters
for entrainment, and it is not at all clear that the standard
picture is correct. In fact, recent work casts some doubt 
on its validity. To see this, we follow Langlois et al \cite{LSC98}
and quantify the coupling timescale in the following way:
\beq
\tau \sim { 1\over 2} \left( c_r + c_r^{-1} \right) \left( 1 + {I_n 
\over I_p} \right)^{-1} {1 \over \Omega_n}
\label{life}\eeq
where $I_n$ and $I_p$ are the moments of inertia of the neutrons
and the protons, respectively, and $c_r$ is the ``drag to lift ratio''
that depends on the relative rotation rate. This estimate is convenient
because it describes the 
lifetime of a local difference in angular velocity of the two fluids
in terms of a single parameter.
As is immediately clear from (\ref{life}) the timescale is 
short if $c_r$ is roughly of order unity (when the ``drag'' resulting 
from entrainment
is of the same order as the Magnus force that acts on the vortices), 
while it becomes
long if $c_r>>1$ or $c_r <<1$. Now, recent estimates of the 
strength of the pinning of the superfluid vortices to the 
nuclei in the inner crust suggest that $c_r$ is small in this 
region (leading to a relaxation time of $\sim 50$~days) \cite{pbjones}. 
Additionally, a detailed investigation of the scattering
of electrons off of proton vortex clusters \cite{sedr}
in the core suggests that $c_r$ is very 
large there. In other words, it would seem plausible that the 
superfluid components relax towards co-rotation with the 
normal fluid rather slowly---both in the crust and deep in the 
core.  This would indicate 
that differential rotation between the superfluid
neutrons and the ``protons'' can be sustained on 
timescales of days to years.

Clearly, this is a problem where we are far away from a 
final conclusion at this point in time. Fortunately, this 
does not matter for the present discussion. It is clear that
even if the coupling between the two fluids is
strong, differential rotation should prevail on a time-scale
orders of magnitude larger than the  dynamical one ($\sim$~ms).
Further evidence for this may follow from observed temperatures 
of millisecond pulsars. We recall attempts to explain the observations
in terms of frictional heating due to differential rotation 
between the superfluid and the crust \cite{shibaz}. 
This discussion typically suggests that roughly
1\% of the total moment of inertia (eg. the superfluid
in the inner crust) must rotate differently from the 
rest of the star, in such a way that 
$10^{-2} \mbox{ rad/s} \le | \Omega_n - \Omega_p | \le 10 \mbox{ rad/s}$.
This would lead to a significant internal heating that would 
be relevant after the neutrino cooling era.

It is also worth mentioning that one should be careful 
before concluding that the 
coupling between the two fluids will necessarily lead to 
a co-rotating system. As argued by Langlois et al \cite{LSC98}, 
this is unlikely to happen in cases when an external torque is 
acting on the system. 
Then the difference $\Omega_n - \Omega_\p$ will stabilize at
a value that represents a balance between the acting torque
and angular momentum redistribution in the fluid. 
Since the external 
torque acts on the crust which then exerts a drag on the 
superfluid, 
we would typically expect $\Omega_\n > \Omega_\p$ in an isolated 
neutron star that spins down due to magnetic dipole braking, while
$\Omega_\n < \Omega_\p$ in a neutron star accreting material
from a binary companion. In other words, one would expect the 
superfluid neutron to spin faster than the charged components
in an isolated radiopulsar that is spinning down due to 
magnetic dipole braking, while the opposite is the case for accreting
neutron stars in (say) the Low-Mass X-ray binaries. 
Given that the latter may actually generate detectable gravitational
waves, see \cite{bil,aks,BC99}, this is an interesting idea.
In fact, it may well turn out that internal differential
rotation needs to be accounted for in the construction 
of accurate theoretical templates for the gravitational
waves from (say) Sco-X1.   
 
During the past decades an accurate machinery for determining
rapidly rotating neutron star models in general relativity
has been developed, see \cite{FI,niksterg} for reviews. 
However, as should be clear from the 
above discussion, truly
realistic neutron star calculations must account for
features associated with the interior superfluid. 
In particular, realistic 
models for rotating neutron 
stars must allow the two fluids to rotate at different
rates (see \cite{prix} for a study of this problem in the Newtonian context). 
This requires a potentially very complex
extension of all existing work. 
Fortunately, given the recent 
development of a relativistic formalism for describing superfluids 
and the accurate numerical codes used to construct rotating
single-fluid stars, many of the tools required to approach this
problem are in hand. Hence, it is timely to initiate a
program aimed at improving our understanding of rotating 
relativistic superfluid stars. 

As a first step towards the solution of this problem, 
the present paper concerns a formalism 
for modeling slowly rotating superfluid neutron 
stars.  Our main aim here is to develop the mathematical
framework and explore how the extra degrees of freedom associated 
with the superfluid affect slowly rotating  
neutron star configurations. The derived
formalism  can then serve as the starting point 
for relativistic studies of pulsar
glitches, or spin-up of neutron stars due to accretion of matter.
Furthermore, the present work provides the first step towards analyzing the 
oscillations of slowly rotating relativistic superfluid neutron stars.
In this sense, the present work is the next logical step in the 
program initiated by Comer et al \cite{CLL}.  Eventually, we aim to 
carry over to general relativity the pioneering 
calculations of Lindblom and Mendell \cite{M,ML,LM1} that were done for 
Newtonian stars. 

There are several reasons why this 
a very important target. First of all, several studies 
of the oscillations of non-rotating superfluid neutron stars 
\cite{Ep,LM1,ulee,CLL} have shown that the new degrees of freedom that 
result because 
the neutrons flow independently of the protons lead to new
sets of oscillation modes (in addition to the familiar f, p and g-modes). 
One interesting question for follow-up studies
of the present work
concerns the effects of rotation on these superfluid modes.
Another motivation for studying oscillations of slowly
rotating superfluid configurations is provided by the gravitational-wave
driven instability of the so-called r-modes \cite{A1,FM}.
A study of the r-modes in a relativistic superfluid star would be 
a natural generalisation of recent work of Lockitch, Andersson and Friedman
\cite{laf00}
and would provide further insights into the details of 
such modes in realistic neutron stars.  
This is a crucial issue given the possibility that the
gravitational waves from the r-modes are potentially
detectable by LIGO~II \cite{owen,BC99}.
One can
speculate that the new degrees of freedom 
associated with the superfluid may lead to new sets of  
r-modes (especially if the neutrons and the protons are not 
corotating).  But even if this possibility is not realized, 
it is important to pursue an analysis of the r-modes 
because of the potentially crucial dissipation due to mutual friction.  
This issue has so far only been addressed in Newtonian gravity \cite{LM00}, 
and there is an obvious need for a detailed relativistic analysis.
 
As already mentioned above, we describe a superfluid neutron star
as a two-fluid system.  Our equations
include vortices implicitly (i.e. the vorticity of the two fluids are not 
zero), and as argued by Comer et al \cite{CLL} should approximate quite 
well the rotational properties.  What are neglected are such physical 
effects as tension along the vortices, and modes that travel along 
individual vortices or are associated with the entire array (e.g. 
Tkachenko waves \cite{TT}).  We do not assume a priori that 
the neutrons rotate at the same rate as the protons in deriving
our equations, nor will chemical 
equilibrium be assumed. This means that we 
can clearly distinguish how the individual rotation rates of 
the neutron and protons (which represent independent degrees of 
freedom), and chemical equilibrium affect the rotational equilibrium.  
Perhaps most importantly, we wish to gain insight into how entrainment 
affects the structure (and, eventually, the dynamics) of neutron stars.
This clearly requires different rotation rates for the neutrons
and the protons since entrainment is no longer relevant when  
the two fluids flow together. 

Our results are presented in the following fashion:  In  
Section II some background material will be presented that determines 
the most general, exact form for the metric and matter variables that 
is allowed by the underlying symmetries and asymptotic conditions.  It 
will be shown that the matter field equations can be solved 
analytically for the 
case of rigid rotations of both the neutron and proton fluids.  
Section III introduces the slow-rotation approximation and applies 
it to the superfluid and Einstein field equations. 
In Section IV we discuss the numerical solutions to the field 
equations, and discuss results for a simple model equation of state. 
 Finally, we discuss issues regarding ``physical'' sequences of stars, 
such as baryon number conservation, the Kepler limit and chemical 
equilibrium in Section V. Our  concluding remarks 
are offered in Section VI.  Some further details that are useful to the 
derivations discussed in the main text are presented in  Appendix~I.  
Except where explicitly noted we use units such that $G = c = 1$.

\section{Axisymmetric, Stationary, and Asymptotically Flat 
Spacetimes}

The study of 
axisymmetric, stationary, and asymptotically flat spacetimes has by now 
a long history.
This means that there is a considerable literature on the subject, and
we refer the reader to  \cite{FI,niksterg} for detailed reviews. 
Our modest constribution in this paper is to construct rotating 
superfluid stellar models within the slow-rotation approximation.
Key references are Comer et al \cite{CLL} and Langlois 
et al \cite{LSC98} for the superfluid formalism, Bonazzola et al 
\cite{BGSM} for how to set up axisymmetric, stationary, and asymptotically 
flat spacetimes, and the work of Hartle and collaborators 
\cite{H1,HT68} for the slow-rotation approximation in relativity. 

\subsection{The Conditions of Axisymmetry, 
Stationarity, and Asymptotic Flatness }

Axisymmetric, stationary, and 
asymptotically flat spacetimes are defined in the following way
\cite{BGSM}: (i) 
There exists a Killing vector, to be denoted here as $t^{\mu}$, that 
is timelike at spatial infinity; (ii) there exists a Killing vector, to 
be denoted here as $\phi^{\mu}$, that vanishes on a timelike two-surface 
(called the axis of symmetry), is spacelike everywhere else, and whose 
orbits are closed curves; and (iii) asymptotic flatness means the scalar 
products $t_{\nu} t^{\nu}$, $\phi_{\nu} \phi^{\nu}$ and $t_{\nu} 
\phi^{\nu}$ go to, respectively, $- 1$, $+ \infty$, and $0$ at spatial 
infinity.  We will likewise impose the so-called ``circularity condition'' 
on the stress-energy tensor $T^{\mu}_{\nu}$, which means that 
\beq
    T^{\mu}_{\nu} t^{\nu} = \alpha t^{\mu} + \beta \phi^{\mu} \qquad , 
            \qquad
    T^{\mu}_{\nu} \phi^{\nu} = \lambda t^{\mu} + \sigma \phi^{\mu} \ .
\eeq

Carter \cite{BC1} has shown that assumptions (i), (ii), and (iii) above 
imply that the Killing vectors commute.  It thus follows that a 
coordinate system, $x^0 = t$, $x^1$, $x^2$, and $x^3 = \phi$, exists for 
which
\beq
    t^{\mu} = (1,0,0,0) \qquad , \qquad \phi^{\mu} = (0,0,0,1) \ .
\eeq
Carter \cite{BC2} has also shown that the ``circularity condition'' 
implies that the metric tensor can be reduced to 
\begin{eqnarray}
    g_{\mu \nu} {\rm d}x^{\mu} {\rm d}x^{\nu} &=& - N^2 {\rm d}t^2 + 
    g_{\phi \phi} \left({\rm d}\phi - N^{\phi} {\rm d}t\right)^2 + 
    g_{1 1} \left({\rm d}x^1\right)^2 + 2 g_{1 2} {\rm d}x^1 {\rm d}x^2
    \cr 
    &&+ g_{2 2} \left({\rm d}x^2\right)^2
\end{eqnarray}
where each of the components depends on only the coordinates $x^1$ and 
$x^2$.  By choosing so-called orthogonal coordinates, one may further 
simplify the metric so that $g_{1 2} = 0$ without any loss of 
generality.

There are a few different orthogonal coordinate systems that have been 
used to study the types of
spacetimes under discussion here.  Bonazolla et al 
\cite{BGSM} discuss several examples that have been used for exact 
calculations, and for studies using the slow-rotation approximation.  
For the present discussion, we will use orthogonal coordinates $x^1 = 
\tilde{r}$ and $x^2 = \theta$ common to slow-rotation studies so that 
the metric is
\begin{eqnarray}
    g_{\mu \nu} {\rm d}x^{\mu} {\rm d}x^{\nu} &=& -\left(N^2 - 
      {\rm sin}^2\theta K \left[N^{\phi}\right]^2\right){\rm d}t^2 +  
      V {\rm d} \tilde{r}^2 - 2 {\rm sin}^2\theta K N^{\phi} {\rm d}t~ 
      {\rm d} \phi \cr 
    && \cr
    &&+ K \left({\rm d}\theta^2 + {\rm sin}^2\theta {\rm d} \phi^2
      \right) \ . \label{finmetric}
\end{eqnarray}

\subsection{Axisymmetric, Stationary, and Asymptotically Flat General 
Relativistic Superfluid Neutron Stars}

We want to see what the various conditions discussed above 
imply for a superfluid neutron star.
Thus we first  recall the formalism that has been  used to model 
general relativistic superfluid neutron stars \cite{CLL,LSC98}.  
The central quantity is the so-called ``master'' function $\Lambda$, 
which is a function of the three scalars $\n^2 = - \n_{\rho} \n^{\rho}$, 
$\p^2 = - \p_{\rho} \p^{\rho}$, and $x^2 = - \p_{\rho} \n^{\rho}$ that 
are formed from $\n^{\mu}$, the conserved neutron number density current, 
and $\p^{\mu}$, the conserved proton number density current.
The master function is such that $-\Lambda$ corresponds to 
the total thermodynamic energy density.

A general variation (that keeps the spacetime metric fixed) of 
$\Lambda(\n^2,\p^2,x^2)$ with respect to the independent vectors 
$\n^{\mu}$ and $\p^{\mu}$ takes the form
\beq
     \d \Lambda = \mu_\rho \d \n^\rho + \chi_\rho \d \p^\rho \ ,
\eeq
where 
\beq
     \mu_{\mu} = \B \n_{\mu} + \A \p_{\mu} \quad , \quad
     \chi_{\mu} = \C \p_{\mu} + \A \n_{\mu} \ , \label{muchidef}
\eeq
and
\beq
   \A = - {\partial \Lambda \over \partial x^2} \qquad , \qquad \B = 
        - 2 {\partial \Lambda \over \partial \n^2} \qquad , \qquad
   \C = - 2 {\partial \Lambda \over \partial \p^2} \ . \label{coef1}
\eeq
The covectors $\mu_{\mu}$ and $\chi_{\mu}$ are dynamically, and 
thermodynamically, conjugate to $\n^{\mu}$ and $\p^{\mu}$, and their 
magnitudes are, respectively, the chemical potentials of the neutrons 
and the protons.  The two covectors also make manifest the entrainment 
effect since it is seen explicitly how the momentum of one constituent 
carries along some mass current of the other constituent (for example, 
$\mu_{\mu}$ is a linear combination of $\n^{\mu}$ 
and $\p^{\mu}$).  We also see that there is no entrainment if the 
master function $\Lambda$ is independent of $x^2$ (because then 
${\cal A}=0$).

The stress-energy tensor is given by 
\beq
     T^{\mu}_{\nu} = \Psi \delta^{\mu}_{\nu} + \p^{\mu} \chi_{\nu} 
                   + \n^{\mu} \mu_{\nu} \ , \label{seten} 
\eeq
where  the generalized pressure $\Psi$ is given by  
\beq
     \Psi = \Lambda - \n^{\rho} \mu_{\rho} - \p^{\rho} \chi_{\rho} \ .
\eeq
The equations of motion consist of two conservation equations,
\beq
  \nabla_{\mu} \n^{\mu} = 0 \qquad , \qquad \nabla_{\mu} \p^{\mu} = 0 
             \ , \label{coneq}
\eeq
and two Euler type equations, which can be conveniently written in 
the compact form 
\beq
     \n^{\mu} \nabla_{[\mu} \mu_{\nu]} = 0 \qquad , \qquad \p^{\mu} 
          \nabla_{[\mu} \chi_{\nu]} = 0 \ . \label{eueqn} 
\eeq
When all four equations are satisfied then it is automatically true that 
$\nabla_{\mu} T^{\mu}_{\nu} = 0$.  

It is convenient for what follows to rewrite each of  
$\n^{\mu}$ and $\p^{\mu}$ as a product of a magnitude with a 
unit timelike vector in such a way that
\beq
     \n^{\mu} = \n u^{\mu} \quad , \quad \p^{\mu} = \p v^{\mu} \ ,
\eeq
where $u^{\mu} u_{\mu} = -1$ and $v^{\mu} v_{\mu} = -1$.  One can 
easily show that the circularity condition, when applied to the 
stress-energy tensor in (\ref{seten}), leads to the following forms 
for the unit timelike vectors $u^{\mu}$ and $v^{\mu}$ (using the 
form of the metric given in (\ref{finmetric})):
\beq
    u^{\mu} = {t^{\mu} + \Omega_{\n} \phi^{\mu} \over 
              \sqrt{N^2 - {\rm sin}^2 \theta K \left(N^{\phi} - 
              \Omega_{\n}\right)^2}} \qquad , \qquad
    v^{\mu} = {t^{\mu} + \Omega_{\p} \phi^{\mu} \over 
              \sqrt{N^2 - {\rm sin}^2 \theta K \left(N^{\phi} - 
              \Omega_{\p}\right)^2}} \ ,
\eeq
where $\Omega_{\n}$ and $\Omega_{\p}$ are the angular velocities of 
the neutrons and protons, respectively.  

With the appropriate decomposition of the unit vectors established, the 
remaining matter variables can be similarly decomposed, and the matter field 
equations can be analyzed.  It turns out that the two conservation 
equations (\ref{coneq}) are automatically satisfied.  
The other matter field equations 
are the two Euler relations (\ref{eueqn}), which can be shown to reduce to
\beq
    \partial_{\nu}\left(t^{\mu} \mu_{\mu}\right) + \Omega_{\n} 
    \partial_{\nu}\left(\phi^{\mu} \mu_{\mu}\right) = 0 \qquad , \qquad
    \partial_{\nu}\left(t^{\mu} \chi_{\mu}\right) + \Omega_{\p} 
    \partial_{\nu}\left(\phi^{\mu} \chi_{\mu}\right) = 0 
\eeq
where
\begin{eqnarray}
   t^{\mu} \mu_{\mu} &=& - {N \B \n \left(1 - {\rm sin}\theta \sqrt{K} 
           N^{\phi} \omega_{\n}/N\right) \over \sqrt{1 - \omega^2_{\n}}} - 
           {N \A \p \left(1 - {\rm sin}\theta \sqrt{K} N^{\phi} \omega_{\p}
           /N\right) \over \sqrt{1 - \omega^2_{\p}}} \ , \cr 
   t^{\mu} \chi_{\mu} &=& - {N \C \p \left(1 - {\rm sin}\theta \sqrt{K} 
           N^{\phi} \omega_{\p}/N\right) \over \sqrt{1 - \omega^2_{\p}}} - 
           {N \A \n \left(1 - {\rm sin}\theta \sqrt{K} N^{\phi} \omega_{\n}
           /N\right) \over \sqrt{1 - \omega^2_{\n}}} \ , \cr
   \phi^{\mu} \mu_{\mu} &=& - {\B \n {\rm sin}\theta \sqrt{K} \omega_{\n} 
           \over \sqrt{1 - \omega^2_{\n}}} - {\A \p {\rm sin}\theta 
           \sqrt{K} \omega_{\p} \over \sqrt{1 - \omega^2_{\p}}} \ , \cr
   \phi^{\mu} \chi_{\mu} &=& - {\C \p {\rm sin}\theta \sqrt{K} \omega_{\p} 
           \over \sqrt{1 - \omega^2_{\p}}} - {\A \n {\rm sin}\theta 
           \sqrt{K} \omega_{\n} \over \sqrt{1 - \omega^2_{\n}}}  
           \ , \label{proj} 
\end{eqnarray}
and 
\beq
    \omega_{\n} \equiv {{\rm sin}\theta \sqrt{K} \left(N^{\phi} - 
                \Omega_{\n}\right) \over N} 
    \quad , \quad 
    \omega_{\p} \equiv {{\rm sin}\theta \sqrt{K} \left(N^{\phi} - 
                \Omega_{\p}\right) \over N} \ .
\eeq
We will focus most of our attention on the case of rigid rotation. 
Then  $\Omega_{\n}$ and $\Omega_{\p}$ are both constants and the Euler 
equations lead to the following first integrals of the motion for 
the neutron and protons, respectively:
\begin{eqnarray}
    \mu_c &=& N \left(\B \n \sqrt{1 - \omega^2_{\n}} + {\A \p 
            \left[1 - \omega_{\n} \omega_{\p}\right] \over \sqrt{1 
            - \omega^2_{\p}}}\right) \ , \cr 
            && \cr
    \chi_c &=& N \left(\C \p \sqrt{1 - \omega^2_{\p}} + {\A \n 
            \left[1 - \omega_{\n} \omega_{\p}\right] \over \sqrt{1 
            - \omega^2_{\n}}}\right) \ , \label{1st_int}
\end{eqnarray}
where $\mu_c$ and $\chi_c$ are both constants.  When the rotation is 
set to zero for both fluids, we retain the first 
integrals obtained by Comer et al \cite{CLL}.  

Given the above results
the superfluid field equations have been completely solved.
However, we still need to determine the corresponding metric 
functions. To do this we must consider the Einstein field equations.  
Unfortunately, these are sufficiently complicated that the same level 
of progress in solving them can not be achieved. That is, we cannot 
write down the solution in closed form.  Such a solution was obviously 
not expected since the Einstein equation requires a numerical solution
already in the one-fluid case. 

The equations we have discussed 
so far do not include any approximations (apart from the 
particular two-fluid model for the superfluid). This means that they 
will be relevant irrespective of the stars rotation rate.
Thus, they could in principle be used as a basis for a 
numerical solution for rapidly rotating stars, following
in the footsteps of \cite{BGSM}. However, we want to 
explore the new degrees of freedom that come into 
play when we consider a two-fluid system. To do this, it seems
natural to first consider the problem in the slow-rotation 
approximation where the equations are somewhat more 
transparent and we can make further ``analytical''
progress. We will return to the problem of rapidly
spinning stars in the future. 

In the following Section 
we will introduce the slow-rotation approximation 
and derive the relevant field equations. These will subsequently be 
integrated numerically, with sample results being discussed in Sections 
IV and V. 

\section{The Slow-Rotation Approximation}

In order for the slow-rotation approximation
to be valid,
the angular velocities must be small enough that the fractional changes in 
pressure, energy density, and gravitational field due to the rotation 
are all relatively small.  When applied to our system the 
approximation can be translated into the inequalities, cf. \cite{H1}
\beq
    \Omega^2_{\n}~{\rm or}~\Omega^2_{\p}~{\rm or}~\Omega_{\n} 
    \Omega_{\p}~<<~\left({c \over R}\right)^2 {G M \over R c^2} \ ,
\eeq  
where the speed of light $c$ and Newton's constant $G$ have been 
restored, and $R$ and $M$ are the radius and mass, respectively, of the 
non-rotating configuration.  Since $G M/c^2 R < 1$, the inequalities 
also imply 
\beq
    \Omega_{\n} R << c \qquad {\rm and} \qquad \Omega_{\p} R << c \ .
\eeq
These conditions indicate that the slow-rotation approximation 
ought to be useful for most astrophysical neutron stars. 

For example, we can compare the above conditions, e.g. that
\beq
[\Omega_n,\Omega_p, \sqrt{\Omega_n\Omega_p}] 
<< \sqrt{ G M\over R^3} \approx 11500  
\left( { M\over M_\odot} \right)^{1/2} 
\left( {10~\mbox{km} \over R} \right)^{3/2}
\ \mbox{ s}^{-1} \ ,
\label{limit}\eeq
to the 
empirical estimate for the Kepler frequency (i.e. the 
rotation rate at which mass-shedding sets in at the 
equator) that has been deduced from calculations using realistic supranuclear
equations of state~\cite{FI}:
\beq
\Omega_K \approx 7600-7700 \left( { M\over M_\odot} \right)^{1/2} 
\left( {10~\mbox{km} \over R} \right)^{3/2}
\ \mbox{ s}^{-1} \ . \label{empire}
\eeq 
 We should also 
recall that the fastest observed pulsar rotates with a period of
1.56~ms. This corresponds to $\Omega \approx 4000~\mbox{s}^{-1}$, i.e.
roughly half the Kepler rate.
In view of these examples, it is not too surprising that
calculations for one-fluid models have shown the slow-rotation
approximation to be accurate to within (say) ten 
percent for stars spinning at the
mass-shedding limit.

Before combining the slow-rotation approximation with our superfluid
formalism, it is worthwhile emphasizing two key points 
that follow since we are dealing with two distinct fluids:
(i) If only one of the rotational directions is reversed, then the 
physical configuration of the star should change; but (ii) if both 
rotational directions are reversed, then the physical 
configuration should not change. As it turns out,
the only quantities that contain terms linear in the angular 
velocities are the metric coefficient $N^{\phi}$, that represents
the dragging of inertial frames, and the fluid four-velocities 
$u^{\mu}$ and $v^{\mu}$.  Thus all other effects due to rotation enter 
only at the second-order in the angular velocities.  The 
calculations performed here, then, will keep terms through second-order.
  
\subsection{Slow Rotation Expansion of the Metric and Matter Variables}

Not very surprisingly, the equations that determine the metric
variables in the slow-rotation approximation for our two-fluid
system are similar to the one-fluid ones derived by Hartle
more than 30 years ago \cite{H1}. Of course, we want to allow 
the two fluids to rotate at different rates so there are some 
conceptual differences associated with the fluid variables. 
Still, it is natural to follow Hartle \cite{H1} and 
expand the five independent metric coefficients as

\begin{eqnarray}
    N &=& e^{\nu(\tilde{r})/2} \left(1 + h(\tilde{r},\theta)\right) 
          \ , \cr
       && \cr 
    V &=& e^{\lambda(\tilde{r})} \left(1 + 2 v(\tilde{r},\theta)\right) 
          \ , \cr 
       && \cr
    K &=& \tilde{r}^2 (1 + 2 k(\tilde{r},\theta)) \ , \cr
       && \cr
    N^{\phi} &=& \omega(\tilde{r},\theta) \ ,
\end{eqnarray}
where it is to be understood that the terms $h$, $v$ and $k$ are each 
second-order in the angular velocites whereas the frame-dragging 
$\omega$ is a first-order quantity.  Initially we will assume that the
two fluids rotate rigidly, which means that there is no need to write 
explicitly similar expansions for the fluid velocities.  Thus, for the 
fluid we need only define the slow-rotation expansion for the neutron and 
proton number densities $\n$ and $\p$, respectively:
\beq
    \n = \n_{\rm o}(\tilde{r}) \left(1 + \eta(\tilde{r},\theta)
         \right) \qquad , \qquad 
    \p = \p_{\rm o} (\tilde{r}) \left(1 + \Phi(\tilde{r},\theta) 
         \right) \ ,
\eeq
where the terms $\eta$ and $\Phi$ are understood to be second-order in 
the angular velocities and we have introduced the convention (to 
be used throughout the paper) that terms with an ``${\rm o}$'' 
subscript are either 
contributions from the non-rotating background or quantities that 
are evaluated on the non-rotating background (e.g. $x^2_{\rm o} = 
\n_{\rm o} \p_{\rm o}$ etc.).  The expansions of the remaining fluid 
variables (such as the velocities, the stress-energy tensor components, 
etc.) can all be obtained in terms of the metric and particle number 
density relationships written above (some of the more useful results 
are presented in Appendix~I).  

A non-rotating background configuration is specified once $\n_{\rm o}$, 
$\p_{\rm o}$, $\nu$ and $\lambda$ are known.  The two background metric 
components are obtained as solutions to
\beq
  \lambda^{\prime} = {1 - e^{\lambda} \over \tilde{r}} - 8 \pi \tilde{r} 
                     e^{\lambda} \Lambda_{\rm o} 
                     \quad , \quad
    \nu^{\prime} = - {1 - e^{\lambda} \over \tilde{r}} + 8 \pi \tilde{r} 
                   e^{\lambda} \Psi_{\rm o}
                   \  \label{bckgrnd}
\eeq
while the neutron and proton number densities are 
obtained from (see \cite{CLL} 
for a complete discussion)
\begin{eqnarray}
    0 &=& \left.\A^0_0\right|_{\rm o} \p_{\rm o}^{\prime} + 
          \left.\B^0_0\right|_{\rm o} \n_{\rm o}^{\prime} + {1 \over 2} 
           (B_{\rm o} \n_{\rm o} + A_{\rm o} \p_{\rm o}) 
           \nu^{\prime} \ , \cr
       && \cr
    0 &=& \left.\C^0_0\right|_{\rm o} \p_{\rm o}^{\prime} + 
          \left.\A^0_0\right|_{\rm o} \n_{\rm o}^{\prime} + {1 \over 2} 
          (A_{\rm o} \n_{\rm o} + C_{\rm o} \p_{\rm o}) 
           \nu^{\prime} \ . \label{bgndfl}
\end{eqnarray}
where
\begin{eqnarray}
\A_0^0 &=& \A + 2 {\partial \B \over \partial \p^2} \n \p + 2 
          {\partial \A \over \partial \n^2} \n^2 + 2 {\partial \A 
          \over \partial \p^2} \p^2 + {\partial \A \over \partial 
          x^2} \p \n \ , \cr
        && \cr
\B_0^0 &=& \B + 2 {\partial \B \over \partial \n^2} \n^2 + 4 
          {\partial \A \over \partial \n^2} \n \p + {\partial \A 
          \over \partial x^2} \p^2 , \cr
        && \cr
\C_0^0 &=& \C + 2 {\partial \C \over\partial \p^2} \p^2 + 4 {\partial 
           \A \over \partial \p^2} \n \p + {\partial A \over \partial 
           x^2} \n^2 \ . 
\end{eqnarray}
Regularity of the geometry at the center of the star requires that 
$\lambda(0)$, $\lambda^{\prime}(0)$, $\nu^{\prime}(0)$, 
$\n_{\rm o}^{\prime}(0)$, and $\p_{\rm o}^{\prime}(0)$ all vanish.  The 
surface of the star is the value $R$ of the radial variable for which 
$\Psi_{\rm o}(R) = 0$.  Finally, the total mass $M$ of the configuration 
is given by
\beq
    M = - 4 \pi \int_0^R \tilde{r}^2 \Lambda_{\rm o}(\tilde{r}) {\rm d} 
        \tilde{r} \ .
\eeq

The equations that determine the rotational features are obtained by taking 
the expansions given above and putting them into the full superfluid and 
Einstein field equations, but keeping only terms up to second-order in 
the rotational velocities.  Even with the slow-rotation 
approximation, this new set of 
equations represents a two-dimensional problem.  Fortunately, the dimension 
represented by the $\theta$ coordinate can be successfully taken care of by 
an expansion in Legendre polynomials $P_l(\cos \theta)$, and  then
it can be shown that we need only consider the $l = 0, 1, 2$ components 
(the argument is completely analogous to that of Hartle \cite{H1}).  
The net result is that the metric 
corrections $h$, $v$, and $k$ can be written as
\begin{eqnarray}
    h &=& h_0(\tilde{r}) + h_2(\tilde{r}) P_2({\cos}\theta) \ , \cr
       && \cr
    v &=& v_0(\tilde{r}) + v_2(\tilde{r}) P_2({\cos}\theta) \ , \cr
       && \cr
    k &=& k_0(\tilde{r}) + k_2(\tilde{r}) P_2({\cos}\theta) \ ,
\end{eqnarray}
where $P_2({\cos}\theta) = (3 {\cos}^2\theta - 1)/2$.  Futhermore, 
Hartle \cite{H1} has shown that a coordinate transformation can be 
imposed whose sole effect is that $k_0(\tilde{r})$ vanishes.  
As for the remaining metric piece $\omega$, it can be 
shown  that it is independent of $\theta$, exactly as in the 
one fluid case, so that we can write $\omega = \omega(\tilde{r})$.  It 
will be useful for what follows to define the quantities
\beq
    \tilde{L}_{\n} = \omega - \Omega_{\n} \quad , \quad 
    \tilde{L}_{\p} = \omega - \Omega_{\p} \ .
\eeq
Up to an overall minus sign, these represent the two rotation frequencies 
as measured by a local zero-angular momentum observer.

The remaining two matter fields $\eta$ and $\Phi$ are written as
\beq
    \eta = \eta_0(\tilde{r}) + \eta_2(\tilde{r}) P_2({\cos}\theta) 
             \quad , \quad
    \Phi = \Phi_0(\tilde{r}) + \Phi_2(\tilde{r}) P_2({\cos}\theta)  
             \ .
\eeq
It is also convenient at this point to write each 
constant that appears in the first integrals of the Euler equations  
(\ref{eueqn})
as a sum of two terms, one (e.g. $\mu_{\infty}$) which is for 
the non-rotating background and a correction (essentially $\gamma_{\n}$) 
which is 
second-order in the rotational velocity, i.e.
\beq
    \mu_c = \mu_{\infty} \left(1 + \gamma_{\n}\right) \quad , \quad
    \chi_c = \chi_{\infty} \left(1 + \gamma_{\p}\right) \ .
\eeq

To complete the system of equations that determines a slowly
rotating superfluid model, we need the Einstein field equations. 
The field equations that will be used here are $G^0_0 
= 8 \pi T^0_0$, $G^1_1 = 8 \pi T^1_1$, $R_{0 3} = 8 \pi \left(T_{0 3} - 
[1/2] T g_{0 3}\right)$, $R^2_2 - R^3_3 = 8 \pi \left(T^2_2 - T^3_3
\right)$, and $R_{1 2} = 0$.  The only matter field equations that need 
to be considered are the first integrals in (\ref{1st_int}).  
It should be noted that not all of these field equations are independent; 
a proper subset will be extracted in Sec. IV where we discuss the
numerical solutions.  

\begin{figure}[tbh]
\centerline{\epsfysize=5cm \epsfbox{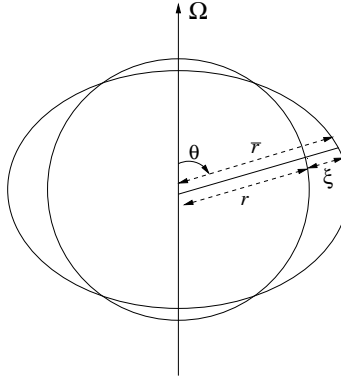}} 
\caption{A schematic illustration of the rotationally distorted
constant density surfaces and the meaning of the coordinate transformation
$\tilde{r}\to r+ \xi$
that is used in the slow-rotation case.}
\label{isos}
\end{figure}

Finally, we need to address a subtle point (discussed in 
detail by Hartle \cite{H1}): Near the surface of the star the 
conditions will be such that, for instance, the ratio of perturbed to 
non-perturbed neutron number density will be ill-defined because the 
non-perturbed background density will vanish.  That this will be the case
is easy to understand 
since the centrifugal force changes the shape of the star. 
To overcome this 
complication, we introduce a 
new radial coordinate $r$  given by
\beq
    \tilde{r} = r + \xi(r,\theta) 
\eeq
such that
\beq
    \Lambda(\tilde{r}(r,\theta),\theta)  = \Lambda_{\rm o}(r) \ .
\eeq
That is, $r$ measures the average distance from the centre
to the rotational surfaces of constant energy density \cite{newcoor}. 
The situation is illustrated schematically in Figure~\ref{isos}.
We then expand the displacement function $\xi$ as
\beq
    \xi = \xi_0(r) + \xi_2(r) P_2(\cos\theta) \ .
\eeq
The shape of the surface of the star can now be obtained as
\beq
    \tilde{R}(\theta) = R + \xi(R,\theta) \ .
\eeq
To linear order in $\xi$ we now  find that
\beq
    \Lambda(r,\theta) = \Lambda_{\rm o}(r)  - \Lambda_{\rm o}^{\prime}(r) 
                        \xi(r,\theta) \ ,
\eeq
where the prime denotes a derivative with respect to $r$.
This and Eq. (\ref{lam_pert}) from Appendix~I imply for $l = 0$ that
\beq
    \mu_{\rm o} \n_{\rm o} \eta_0 + \chi_{\rm o} \p_{\rm o} \Phi_0 + 
    {r^2 \over 3 e^{\nu}} \A_{\rm o} \n_{\rm o} \p_{\rm o} 
    \left(\Omega_{\n} - \Omega_{\p}\right)^2 = 
    \Lambda^{\prime}_{\rm o} \xi_0 \label{coortrans0}
\eeq 
while for $l = 2$ we have
\beq
    \mu_{\rm o} \n_{\rm o} \eta_2 + \chi_{\rm o} \p_{\rm o} \Phi_2 - 
    {r^2 \over 3 e^{\nu}} \A_{\rm o} \n_{\rm o} \p_{\rm o} 
    \left(\Omega_{\n} - \Omega_{\p}\right)^2  = 
    \Lambda^{\prime}_{\rm o} \xi_2 \ . \label{coortrans2}
\eeq

The coordinate transformation affects the ``03'', ``22 - 33'', and ``12'' 
Einstein equations and the first integrals of the Euler equations only 
to the extent that $\tilde{r}$ gets replaced with $r$.  More substantial 
is what happens to the ``00'' and ``11'' Einstein equations.  For these 
one must transform the corrections to the Einstein and stress-energy 
tensors in the old coordinate system, denoted by $\delta G^{\mu}_{\nu}$ 
and $\delta T^{\mu}_{\nu}$, respectively, to the corrections in the new 
coordinate system, denoted by $\Delta G^{\mu}_{\nu}$ and $\Delta 
T^{\mu}_{\nu}$.  The transformations are readily found to be
\beq
  \Delta G^0_0 = \delta G^0_0(r,\theta) + {\partial \left.G^0_0
                   \right|_{\rm o} \over \partial r} \xi(r,\theta) \quad , 
                \quad
  \Delta G^1_1 = \delta G^1_1(r,\theta) + {\partial \left.G^1_1
                   \right|_{\rm o} \over \partial r} \xi(r,\theta)
\eeq
for the Einstein tensor corrections and
\beq
   \Delta T^0_0 = \delta T^0_0(r,\theta) + {\partial \left.T^0_0
                    \right|_{\rm o} \over \partial r} \xi(r,\theta) \quad ,
                  \quad
   \Delta T^1_1 = \delta T^1_1(r,\theta) + {\partial \left.T^1_1
                    \right|_{\rm o} \over \partial r} \xi(r,\theta) \ . 
\eeq
for the stress-energy tensor corrections.

Now, the first integrals of the Euler equations have as their $l = 0$ 
contributions
\begin{eqnarray}
    \gamma_{\n} &=& {\left.\b00\right|_{\rm o} \n_{\rm o} \over 
                    \mu_{\rm o}} \eta_0 + {\left.\a00\right|_{\rm o} 
                    \p_{\rm o} \over \mu_{\rm o}} \Phi_0 + {r^2 \over 3 
                    e^{\nu}} {\p_{\rm o} \over \mu_{\rm o}} \left(
                    \A_{\rm o} + \n_{\rm o} \left.{\partial 
                    \A \over \partial \n}\right|_{\rm o} + \n_{\rm o} 
                    \p_{\rm o} \left.{\partial \A \over \partial x^2}
                    \right|_{\rm o}\right) \left(\Omega_{\n} - \Omega_{\p}
                    \right)^2 - \cr
                 && \cr
                 && {r^2 \over 3 e^{\nu}} \tilde{L}_{\n}^2 + h_0 \ , \cr
                 && \cr
    \gamma_{\p} &=& {\left.\c00\right|_{\rm o} \p_{\rm o} \over 
                    \chi_{\rm o}} \Phi_0 + {\left.\a00\right|_{\rm o} 
                    \n_{\rm o} \over \chi_{\rm o}} \eta_0 + {r^2 \over 3 
                    e^{\nu}} {\n_{\rm o} \over \chi_{\rm o}} 
                    \left(\A_{\rm o} + \p_{\rm o} \left.{\partial 
                    \A \over \partial \p}\right|_{\rm o} + \n_{\rm o} 
                    \p_{\rm o} \left.{\partial \A \over \partial x^2}
                    \right|_{\rm o}\right) \left(\Omega_{\n} - \Omega_{\p}
                    \right)^2 - \cr
                 && \cr
                 && {r^2 \over 3 e^{\nu}} \tilde{L}_{\p}^2 + h_0 \ , 
                    \label{flden0}
\end{eqnarray}
where $\mu_{\rm o} = \B_{\rm o} \n_{\rm o} + \A_{\rm o} \p_{\rm o}$ and 
$\chi_{\rm o} = \C_{\rm o} \p_{\rm o} + \A_{\rm o} \p_{\rm o}$, and for 
$l = 2$
\begin{eqnarray}
    0 &=& {\left.\b00\right|_{\rm o} \n_{\rm o} \over 
                    \mu_{\rm o}} \eta_2 + {\left.\a00\right|_{\rm o} 
                    \p_{\rm o} \over \mu_{\rm o}} \Phi_2 - {r^2 \over 3 
                    e^{\nu}} {\p_{\rm o} \over \mu_{\rm o}} \left(
                    \A_{\rm o} + \n_{\rm o} \left.{\partial 
                    \A \over \partial \n}\right|_{\rm o} + \n_{\rm o} 
                    \p_{\rm o} \left.{\partial \A \over \partial x^2}
                    \right|_{\rm o}\right) \left(\Omega_{\n} - \Omega_{\p}
                    \right)^2 + \cr
                 && \cr
                 && {r^2 \over 3 e^{\nu}} \tilde{L}_{\n}^2 + h_2 \ , \cr
       && \cr
    0 &=& {\left.\c00\right|_{\rm o} \p_{\rm o} \over 
                    \chi_{\rm o}} \Phi_2 + {\left.\a00\right|_{\rm o} 
                    \n_{\rm o} \over \chi_{\rm o}} \eta_2 - {r^2 \over 3 
                    e^{\nu}} {\n_{\rm o} \over \chi_{\rm o}} 
                    \left(\A_{\rm o} + \p_{\rm o} \left.{\partial 
                    \A \over \partial \p}\right|_{\rm o} + \n_{\rm o} 
                    \p_{\rm o} \left.{\partial \A \over \partial x^2}
                    \right|_{\rm o}\right) \left(\Omega_{\n} - \Omega_{\p}
                    \right)^2 + \cr
                 && \cr
                 && {r^2 \over 3 e^{\nu}} \tilde{L}_{\p}^2 + h_2 \ . 
                    \label{flden2}
\end{eqnarray}
The ``03'' Einstein equation yields
\beq
    {1 \over r^4} \left(r^4 e^{- (\lambda + \nu)/2} 
      \tilde{L}_{\n}^{\prime}\right)^{\prime} - 16 \pi e^{(\lambda - 
      \nu)/2} \left(\Psi_{\rm o} - \Lambda_{\rm o}\right) \tilde{L}_{\n} 
      = 16 \pi e^{(\lambda - \nu)/2} \chi_{\rm o} \p_{\rm o} 
      \left(\Omega_{\n} - \Omega_{\p}\right) \ , \label{frmdrg}
\eeq
which is the equation that determines the frame-dragging.  It is formally 
equal to that obtained by Hartle \cite{H1} except for the non-zero 
source term on the right-hand-side. In the particular case when the 
two fluids are co-rotating ($\Omega_n=\Omega_p$) we retain the 
standard result.  

The ``22 - 33'' and ``12'' equations 
only have $l = 2$ contributions and these are, respectively, 
\begin{eqnarray}
    v_2 + h_2 &=& {r^4 \over 6 e^{\nu + \lambda}} 
                  \left(\tilde{L}_{\n}^{\prime}
                  \right)^2 + {8 \pi r^4 \over 3 e^{\nu}} 
                  \left(\Psi_{\rm o} - \Lambda_{\rm o}\right) 
                  \tilde{L}_{\n}^2   \cr
               && + {8 \pi r^4 \over 3 e^{\nu}} \left[\chi_{\rm o} 
                  \p_{\rm o} \left(\Omega_{\n} - \Omega_{\p}\right) 
                  \left(\tilde{L}_{\n} + \tilde{L}_{\p}\right) - 
                  \A_{\rm o} \n_{\rm o} \p_{\rm o} \left(\Omega_{\n} 
                  - \Omega_{\p}\right)^2\right] \label{222}
\end{eqnarray} 
and
\beq
     {1 \over r} \left(v_2 + h_2\right) - \left(k_2 + h_2
     \right)^{\prime} - {\nu^{\prime} \over 2} \left(h_2 - 
     v_2\right) = 0 \ . \label{122}
\eeq
The $l = 0$ part of the ``00'' equation is
\begin{eqnarray}
   0 &=& {16 \pi r^2 \over 3 e^{\nu}} \left[\left(\Psi_{\rm o} - 
         \Lambda_{\rm o}\right) \tilde{L}_{\n}^2 + \chi_{\rm o} \p_{\rm o} 
         \left(\Omega_{\n} - \Omega_{\p}\right) \left(\tilde{L}_{\n} + 
         \tilde{L}_{\p}\right) - \A_{\rm o} \n_{\rm o} \p_{\rm o} \left(
         \Omega_{\n} - \Omega_{\p}\right)^2\right] + \cr
      && \cr
      && 8 \pi \Lambda^{\prime}_{\rm o} \xi_0 - {2 \over r^2} \left({r 
         \over e^{\lambda}} v_0\right)^{\prime} + {r^2 \over 6 e^{\nu + 
         \lambda}} \left(\tilde{L}_{\n}^{\prime}\right)^2 \label{000}
\end{eqnarray}
with the $l = 2$ piece being
\begin{eqnarray}
   0 &=& - {16 \pi r^2 \over 3 e^{\nu}} \left[\left(\Psi_{\rm o} - 
           \Lambda_{\rm o}\right) \tilde{L}_{\n}^2 + \chi_{\rm o} 
           \p_{\rm o} \left(\Omega_{\n} - \Omega_{\p}\right) \left(
           \tilde{L}_{\n} + \tilde{L}_{\p}\right) - \A_{\rm o} \n_{\rm o} 
           \p_{\rm o} \left(\Omega_{\n} - \Omega_{\p}\right)^2\right] + \cr 
      && \cr
      && 8 \pi \Lambda^{\prime}_{\rm o} \xi_2 - {2 \over r^2} \left({r 
         \over e^{\lambda}} v_2\right)^{\prime} - {r^2 \over 6 e^{\nu + 
         \lambda}} \left(\tilde{L}_{\n}^{\prime}\right)^2 + {2 \over 
         e^{\lambda}} \left[k^{\prime \prime}_2 + \left({3 \over r} - 
         {\lambda^{\prime} \over 2}\right) k^{\prime}_2 - {2 e^{\lambda} 
         \over r^2} k_2\right] - {6 \over r^2} v_2 \ . \label{002}
\end{eqnarray}
Finally, the $l = 0$ part of the ``11'' equation is 
\begin{eqnarray}
   && {2 \over r e^{\lambda}} h^{\prime}_0 - {2 \over r e^{\lambda}} 
      \left(\nu^{\prime} + {1 \over r}\right) v_0 + {r^2 \over 6 
      e^{\nu +\lambda}} \left(\tilde{L}_{\n}^{\prime}\right)^2 = 8 \pi 
      \left[\mu_{\rm o} \n_{\rm o} \gamma_{\n} + \chi_{\rm o} \p_{\rm o} 
      \gamma_{\p} - \left(\Psi_{\rm o} - \Lambda_{\rm o}\right) h_0 + 
      \right. \cr
   && \cr
   && \left.{r^2 \over 3 e^{\nu}} \left(\mu_{\rm o} \n_{\rm o} 
      \tilde{L}_{\n}^2 + \chi_{\rm o} \p_{\rm o} \tilde{L}_{\p}^2\right) 
      - {r^2 \over 3 e^{\nu}} \n_{\rm o} \p_{\rm o} \A_{\rm o} \left(
      \Omega_{\n} - \Omega_{\p}\right)^2\right] \label{110}
\end{eqnarray}
while the $l = 2$ result is
\begin{eqnarray}
   &&{2 \over r e^{\lambda}} h^{\prime}_2 - {6 \over r^2} h_2 - 
     {2 \over r e^{\lambda}} \left(\nu^{\prime} + {1 \over r}\right) 
     v_2 + {1 \over e^{\lambda}} \left(\nu^{\prime} + {2 \over r}
     \right) k^{\prime}_2 - {4 \over r^2} k_2 - {r^2 \over 6 
      e^{\nu +\lambda}} \left(\tilde{L}_{\n}^{\prime}\right)^2 = \cr
   && \cr
   && - 8 \pi \left[\left(\Psi_{\rm o} - \Lambda_{\rm o}\right) h_2 + 
      {r^2 \over 3 e^{\nu}} \left(\mu_{\rm o} \n_{\rm o} 
      \tilde{L}_{\n}^2 + \chi_{\rm o} \p_{\rm o} \tilde{L}_{\p}^2\right) 
      - {r^2 \over 3 e^{\nu}} \n_{\rm o} \p_{\rm o} \A_{\rm o} \left(
      \Omega_{\n} - \Omega_{\p}\right)^2\right]  \ . \label{112}
\end{eqnarray}

Before we proceed, 
it is worth emphasising the fact that
 the $l = 0,1,2$ components of the fluid and metric field 
variables decouple. This is in complete analogy with the results
of Hartle for the one-fluid case \cite{H1}.

\subsection{The Exterior Vacuum Solutions}

In the exterior of the star our problem is identical to the one-fluid one.
Hence, we can use the analytic solutions found by Hartle \cite{H1}.
Thus, the vacuum solutions are written
\beq
   \omega(r) = {2 J \over r^3} \ , 
\label{fvac}\eeq
\beq      
h_0(r) = - {\delta M \over r - 2 M} + {J^2 \over r^3 (r - 2 M)} \ ,
\eeq
\beq
v_0(r) = -h_0(r) =  {\delta M \over r - 2 M} -  {J^2 \over 
         r^3 (r - 2 M)} \ , 
\eeq
\begin{eqnarray}
h_2(r) &=& - A \left[{3 \over 2} \left({r \over M}\right)^2 
                 \left(1 - {2 M \over r}\right) {\rm ln} \left(1 - {2 M 
                 \over r}\right) + {(r - M) \left(3 - 6 M/r 
                 - 2 (M/r)^2\right) \over M (1 - 2 M/r)}\right] + \cr
              && \cr
              && {J^2 \over M r^3} \left(1 + {M \over r}\right) \ , 
\end{eqnarray}
\begin{eqnarray}
      k_2(r) &=& A \left[{3 \over 2} \left({r \over M}\right)^2 \left(1 
                 - {2 M^2 \over r^2}\right) {\rm ln} \left(1 
                 - {2 M \over r}\right) + {3 (r - M) - 8 (M/r)^2 \left(r 
                 - M/2\right) \over M (1 - 2 M/r)}\right] - \cr
              && \cr
              && {J^2 \over M r^3} \left(1 + {2 M \over r}\right)\ , 
\end{eqnarray}
\begin{eqnarray}
      v_2(r) &=& A \left[{3 \over 2} \left({r \over M}\right)^2 
                 \left(1 - {2 M \over r}\right) {\rm ln} \left(1 - {2 M 
                 \over r}\right) + {(r - M) \left(3 - 6 M/r 
                 - 2 (M/r)^2\right) \over M (1 - 2 M/r)}\right] - \cr
              && \cr
              && {J^2 \over M r^3} \left(1 - {5 M \over r}\right) \ ,
\end{eqnarray}
where $J$, $\delta M$ and $A$ are constants. $J$  
represents the total angular momentum of the star with respect to 
inertial observers at spatial infinity while $\delta M$ represents the 
change in total mass-energy due to the rotation. 
The final constant $A$ is related to the quadrupole moment
of the star through (cf. Hartle and Thorne \cite{HT68}, although with 
the sign reversed to agree with Laarakkers and Poisson \cite{LP})
\beq
Q = - {8\over 5} A M^3 - {J^2 \over M} \ ,
\label{qpole}\eeq 
and it is determined when we solve the $l=2$ equations in 
section~IV. This then facilitates a comparison with the 
well-known result for the Kerr black hole spacetime. Namely
that
\beq
Q_{\rm Kerr} = - {J^2 \over M} \ .
\eeq 

Assuming that the metric is continuous at the surface of the star,  
it can be shown that 
\begin{eqnarray}
    J &=& - {8 \pi \over 3} \int_0^R r^4 e^{(\lambda - \nu)/2} \left(
        \Psi_{\rm o} - \Lambda_{\rm o}\right) \tilde{L}_{\n} {\rm d} r 
        - {8 \pi \over 3} \int_0^R r^4 e^{(\lambda - \nu)/2} \chi_{\rm o} 
        \p_{\rm o} \left(\Omega_{\n} - \Omega_{\p}\right) {\rm d} r = 
\nonumber \\
&=& - {8 \pi \over 3} \int_0^R {\rm d}r r^4 e^{(\lambda - \nu)/2}
         \left[\mu_{\rm o} \n_{\rm o} \tilde{L}_{\n} + 
\chi_{\rm o} p_{\rm o} \tilde{L}_{\p} \right] \ .
\end{eqnarray}
It would be tempting to interpret the two terms in the integrand 
as representing the angular momentum in the neutrons and the 
protons, respectively. However, as we will show in Section~VB
this is only correct in the absence of entrainment.
We also have
\beq
    \delta M = (R - 2 M) v_0(R) + {J^2 \over R^3} \ .
\eeq
Using these two equations, the dependence of $J$ and $\delta M$ on the central neutron 
and proton number densities can be established. 

\section{Numerical Results}

There are four essential steps to using the slow-rotation formalism 
to produce a model of a general relativistic superfluid 
neutron star: (i) build a nonrotating background configuration,  (ii) 
determine the frame-dragging by 
solving (\ref{frmdrg}) using as input the background 
configuration, (iii) solve the $l = 0$ equations using as input the 
background configuration and the frame-dragging results, and (iv) 
determine the $l = 2$ contributions by solving the relevant
equations in a similar way.   
In this section we
describe our implementation of these steps 
and present some typical numerical results.

\subsection{The Background Configurations}

The nonrotating background configuration is easily constructed
by solving equations (\ref{bckgrnd}) and (\ref{bgndfl}). 
The required parameters are the 
central number densities $\n_{\rm o}(0)$ and $\p_{\rm o}(0)$. 
For any given equation of state these parameters are linked by the
requirement that the background fluid be in chemical equilibrium. As a 
suitably simple model equation of state, we consider the case when the 
two fluids are described by independent polytropes. The master function 
then takes the form \cite{CLL}
\beq
\Lambda(\n^2,\p^2) = -m_\n \n - \sigma_\n \n^{\beta_\n} 
-m_\n \p - \sigma_\p \p^{\beta_\p} 
\eeq
For simplicity we assume that the neutron and the 
proton masses are identical ($=m_\n$). 
It should also be noted that 
we do not, from this point on, include the entrainment effect, even 
though all equations we have derived so far, in principle, accomodates 
it.  The main reason for not building specific numerical solutions with 
entrainment here is that the mere inclusion of two fluids is significant and 
we feel it is crucial 
to first establish that the slow-rotation framework that we have 
developed produces reliable results. Moreover, the implementation
of entrainment requires a considerable amount of further work
as far as the model equation of state is concerned.
In practice, the fact that we are not including entrainment in 
the models discussed in this and the following section
means that all our numerical calculations assume that
${\cal A} = {\cal A}_0^0 = 0$ etc. A future paper will be devoted to a 
model for entrainment and its effect on rotating neutron stars.

Our model equation of state is the same as that used by Comer et al
\cite{CLL}
in the study of quasinormal modes of non-rotating superfluid stars. 
However, we have chosen our parameters differently in order to make
our star somewhat more realistic. Our aim was to create a model star
with mass roughly $1.4M_\odot$, radius $10$~km and a central
proton fraction of roughly 10\%, i.e. what could be considered
typical values for a realistic neutron star. Given such a model
we can compare the numerical results for (say) the 
increase in inertial mass due to the rotation with results
obtained for more realistic supranuclear equations of state and
stars with realistic parameters. 

The particular values we use for the two polytropes are
\begin{eqnarray*}
\sigma_n &=& 0.2  \ , \\
\beta_n &=& 2.3 \ , \\
\sigma_p &=& 2 \ , \\
\beta_p &=& 1.95 \ .
\end{eqnarray*}

\begin{figure}[tbh]
\centerline{\epsfxsize=9cm \epsfbox{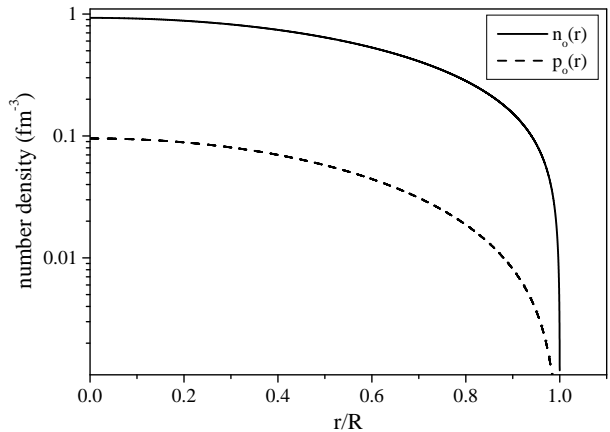}} 
\caption{The neutron and proton number densities as functions
of $r$ for our model star.  }
\label{bkg}
\end{figure}

We assume that the non-rotating model is in chemical equilibrium.
This means that we impose the condition $\mu_{\rm o} = \chi_{\rm o}$ 
throughout the star.
The resultant sequence of stellar models is illustrated in 
figure~\ref{kepseq}. The particular model which will be used to 
illustrate the typical effects of rotation corresponds to a central 
neutron number density $n_{\rm o}(0)= 0.93$~fm$^{-3}$, which 
leads to a total central energy density of 
$-\Lambda_{\rm o}(0) = 1.215$~fm$^{-3}$. The resultant non-rotating
star has $M=1.409M_\odot$ and $R=10.076$~km. 
As is clear from figure~\ref{kepseq} this star is stable
against radial oscillations. 
The neutron and proton density
profiles for our particular model star are shown in figure~\ref{bkg}.

This figure illustrates two 
important things. First of all it is easy to see that we have chosen
our parameters in such a way that the proton fraction at the 
centre of the star is roughly 10\%. Secondly, it should be noted that 
the proton and neutron densities vanish at the same value 
of $r$. In other words, the surface of the star corresponds to 
$n_{\rm o}(R) =p_{\rm o}(R) = 0$. This is, in fact, a 
consequence of our choice to enforce chemical 
equilibrium throughout the star. It obviously means that we are
not at this point considering the effects of regions of the star
that extend beyond the superfluid: the low-density crust region
and the ocean expected to cover the neutron star's surface. 
To work with a simplified stellar model is natural at this 
point, but it is worth pointing out that a formalism 
for relativistic solid crusts was developed by Carter and Quintana
\cite{CQ} a long time ago. 
A potentially very interesting extension of the present work 
concerns the matching of our equations, representing 
the superfluid core, to a slow-rotation version of the 
equations of Priou \cite{denisp}, describing 
the dynamics of solid crusts. 

\subsection{The Frame-Dragging}

Once the background configuration has been determined,
the frame-dragging $\omega$ can be calculated from ({\ref{frmdrg}). 
This calculation is standard, but 
it is notable that the two-fluid results differ from those
of Hartle in that a single integration does not
suffice to determine the frame-dragging at all rates of rotation.
This is simply because the integration of (\ref{frmdrg}) requires 
the central value of $\tilde{L}_n$ as well as both 
the neutron and the proton rotation rates. In practice, we solve a
rescaled version of (\ref{frmdrg}) such that we determine
$\hat{L}_n=\tilde{L}_n/\Omega_p$ given $\hat{L}_n(0)$ and the 
relative rotation rate $\Omega_n/\Omega_p$. This then determines
the frame-dragging for all $\Omega_p$ with a fixed relative rotation.
It is natural to scale the variables with the rotation rate of the 
protons  rather than that of the neutrons since, as we 
discussed in the Introduction, one would expect 
the charged components in a neutron star core to be electromagnetically
coupled to the nuclei in the crust. This locks the two components
(which make up our ``proton'' fluid) together on a very
short timescale. Meanwhile the superfluid neutrons may rotate
differently. Furthermore, as will become clear in Section~V, 
there may be situations where one would need to allow the neutrons
to rotate with $\Omega_n\neq$~constant. 

The solution to the frame-dragging equation (\ref{frmdrg})
is to be such that the interior matches smoothly
onto the known vacuum solution (\ref{fvac}). This means that we must have
\beq
\tilde{L}_n(R) = -\Omega_n + {2J \over R^3} \ .
\label{surfc1}\eeq
We can easily see that $\tilde{L}_n$ and its derivative are 
smooth provided that we have
\beq
\tilde{L}_n(R) =  -\Omega_n - {R \over 3} 
\left. { d\tilde{L}_n \over dr}\right|_{r=R} \ .
\label{ltsurf}\eeq 
Once we find a value for $\tilde{L}_n(0)$ such that (\ref{ltsurf}) 
is satisfied, we have an acceptable solution to the problem,
and we can determine the angular momentum of the configuration
from (\ref{surfc1}). 

Having solved the frame-dragging equation for our model star we find
(not very surprisingly) that the result differs very little 
from the standard one-fluid result as long as the relative rotation 
of the neutrons and the protons is not too extreme.
One typically finds that 
the frame dragging is a smooth function that 
decreases monotonically from the centre of the star towards
the surface. A sample of results are shown in figure~\ref{frame1}.

\begin{figure}[tbh]
\centerline{\epsfxsize=9cm \epsfbox{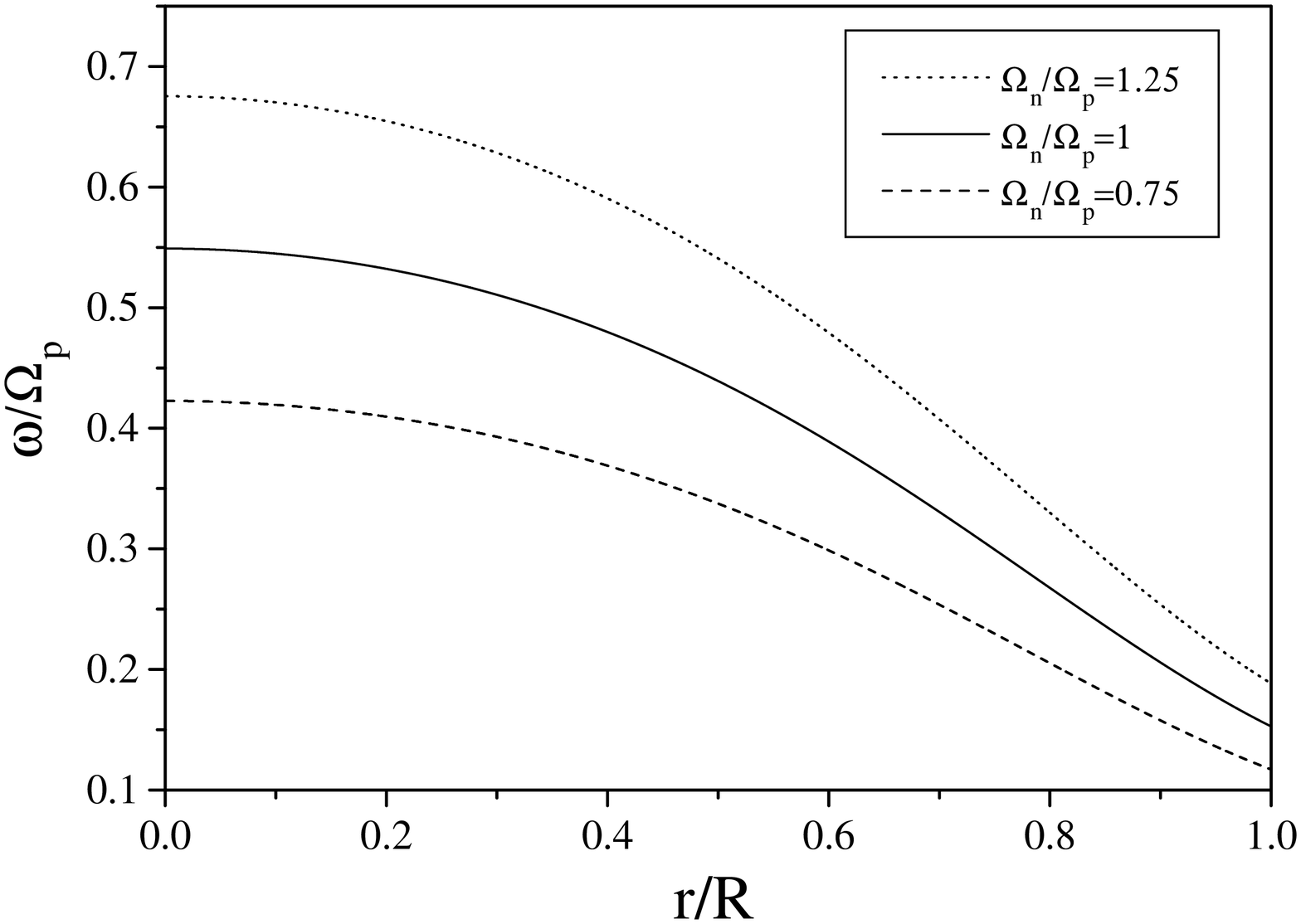}} 
\caption{Typical results for the frame-dragging $\omega$ for superfluid
stars in which the neutrons and the protons rotate at slightly different
rates. }
\label{frame1}
\end{figure}

Even though the outcome may be 
of limited physical significance it is interesting
to consider more extreme situations, such as ones where the neutrons 
are rapidly counter-rotating relative to the protons. An example of such a
case is shown in figure~\ref{frame2}. Here the neutrons rotate backwards
with  respect to the protons in such a way that $\Omega_n =  -0.08\Omega_p$.
This example shows that the two-fluid system allows models where
the frame-dragging  no longer changes
monotonically from the centre to the surface of the star. In the particular
case shown in figure~\ref{frame2} the forward rotation of the 
protons (which make up roughly 10\% of the total number density
at the centre, cf. figure~\ref{bkg}) counteract the frame dragging
induced by the neutrons in the central parts of the star. Meanwhile, the
backwards rotation of the neutrons dominate the frame-dragging in the
surface region.

\begin{figure}[tbh]
\centerline{\epsfxsize=9cm \epsfbox{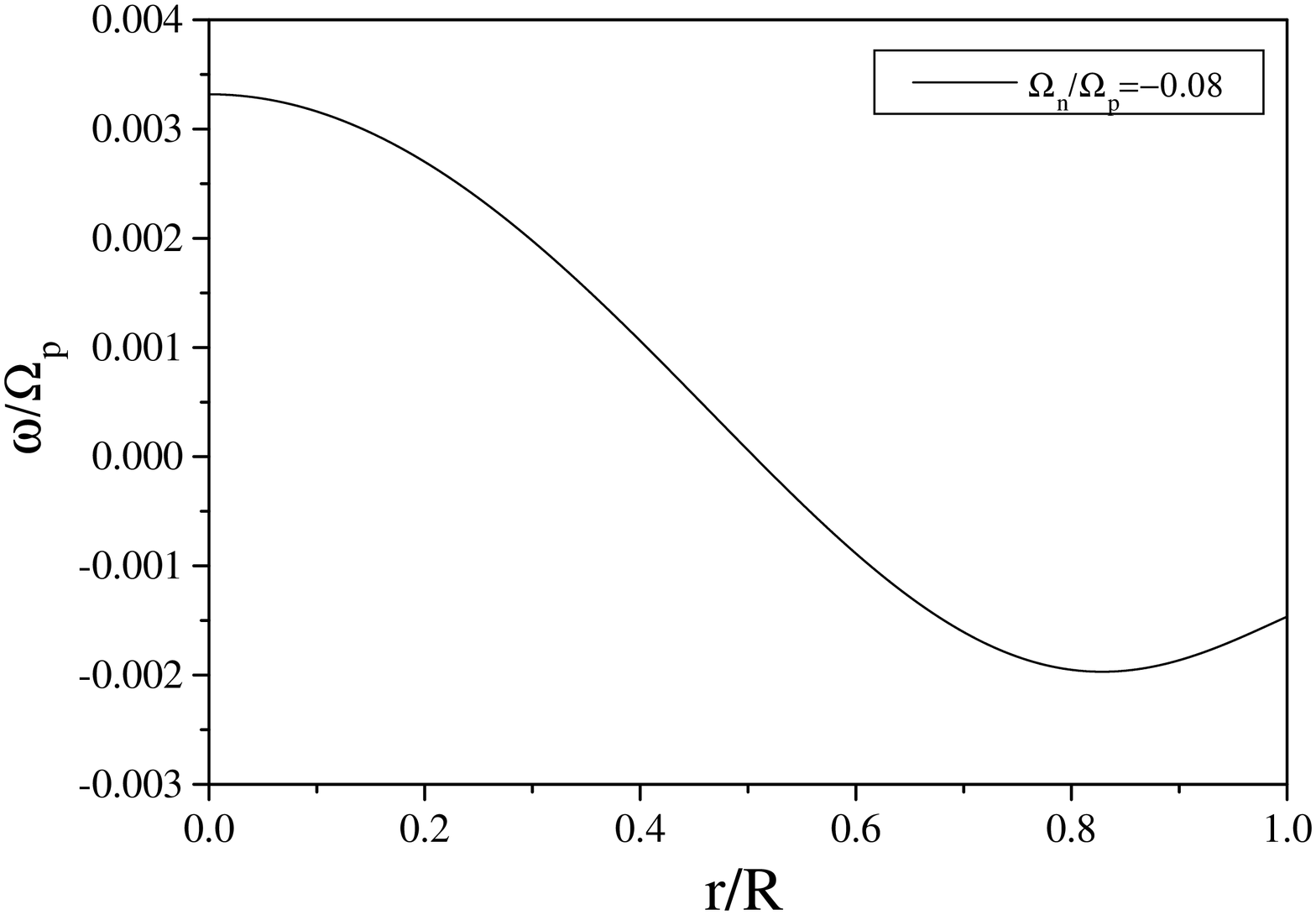}} 
\caption{The frame dragging in an extreme (rather unphysical) situation
where the neutrons and the protons counterrotate in such a way that the 
net frame dragging is ``backwards'' at the surface of the star but 
``forwards'' in the central parts. }
\label{frame2}
\end{figure}

\subsection{The $l = 0$ Results}

Let us now turn to the next phase of the problem, namely the 
$l=0$ equations at the ${\cal O}(\Omega^2)$ level. Thus we consider
Eqns.
(\ref{coortrans0}), (\ref{flden0}), (\ref{000}) and (\ref{110}).
This corresponds to five equations for $\xi_0$, $\eta_0$, $\Phi_0$, 
$h_0$ and $v_0$.  We also need to determine the two integration
constants $\gamma_n$ and $\gamma_p$. In practice we proceed 
as follows: First we introduce a new variable
\beq
m_0 = r e^{-\lambda} v_0 \ .
\eeq
Then we implement the boundary conditions at the centre of the star
by requiring that $m_0(0)=\xi_0(0)=0$ and $h_0(0)=\mbox{ constant}$.
The first two correspond to spacetime being locally flat at the
centre and the requirement that our new coordinate system has the
same origin as the spherical coordinates used to describe
the non-rotating model. It is also worth noticing that the condition
$\xi_0(0)=0$ leads to the central energy density of our rotating
star necessarily being equal to $-\Lambda_{\rm o}(0)$, i.e. the 
central density of the non-rotating star. This is true also 
for the single fluid slow-rotation approximation, cf. \cite{H1}. 

However, in the superfluid case we have an extra degree of freedom.
It is not sufficient to provide the total energy density at
the centre, we need to also provide either $\eta_0(0)$
or $\Phi_0(0)$. These represent the change in the neutron and
proton central density, respectively. The fact that we have 
$\xi_0(0)=0$ relates these two parameters and 
we must have 
\beq
    \left.\left(\mu_{\rm o} \n_{\rm o} \eta_0\right)\right|_{r = 0} + 
    \left.\left(\chi_{\rm o} \p_{\rm o} \Phi_0\right)\right|_{r = 0} 
    = 0 \ .
\eeq 
In other words, we only need to provide $\eta_0(0)$ (say).
In the following we will typically assume that $\eta_0(0)=\Phi_0(0) = 0$,
i.e. that the central proton to neutron ratio
remains unchanged in the rotating star. 
However, this choice is made for convenience
and there are other options that may be
physically more relevant. We discuss such possibilities 
in section~V, but all results presented in this section 
pertain to the case when
$\eta(0)=\Phi_0(0)=0$.

Given the data at the centre of the star we can easily
determine $\gamma_n$ and $\gamma_p$ from (\ref{flden0}).
This leads to 
\begin{eqnarray}
    \gamma_{\n} &=& h_0(0) + \left.\left({\mu_{\rm o} \left.\a00
                  \right|_{\rm o} - \chi_{\rm o} \left.\b00
                  \right|_{\rm o} \over \mu_{\rm o}^2} \p_{\rm o} \Phi_0
                  \right)\right|_{r = 0} , \cr
                 && \cr
    \gamma_{\p} &=& h_0(0) + \left.\left({\mu_{\rm o} \left.\c00
                  \right|_{\rm o} - \chi_{\rm o} \left.\a00
                  \right|_{\rm o} \over \mu_{\rm o} \chi_{\rm o}} 
                  \p_{\rm o} \Phi_0\right)\right|_{r = 0} \ .
\end{eqnarray}

Now we are set to solve the various $l=0$ equations. We do this by 
integrating (\ref{000}) and (\ref{110}) using a standard 
fourth order Runge-Kutta scheme. After each integration step
we determine the relevant values for $\xi_0$, $\eta_0$ and $\Phi_0$
by solving the algebraic equations (\ref{coortrans0}) and (\ref{flden0}). 
At the surface of the star the solutions for 
$h_0$ and $m_0$ must be matched 
to the vacuum solutions given in section~IIIC. 
 Once we find a 
value for $h_0(0)$ such that both $h_0$ and its derivative match 
the vacuum result we have our desired solution to the 
$l=0$ problem.
The matching at the 
surface also determines $\delta M$, the rotationally
induced change in inertial mass of the star. For the particular
stellar model  shown in figure~\ref{bkg} and the case when the 
two fluids are corotating 
we find that
$$
\delta M = 0.091\left({\nu_p \over 1~\mbox{kHz}} \right)^2 M_\odot \ ,
$$
where $\nu_p = \Omega_\p/2\pi$ is the rotation rate of the protons.
As we will discuss in Section~VB the mass-shedding limit for this star 
corresponds to $\nu_p = 1630$~Hz, which means that rotation may increase
the inertial mass of the star by up to roughly 17~\%.
The rotationally induced increase in mass for a sequence
of models is illustrated in figure~\ref{kepseq}. We compare the 
mass of each non-rotating star to the maximum mass (which is 
reached when the star spins at the Kepler limit (see section~V for
a discussion). In the figure we indicate (by an arrow) the 
rotating analogues of the model star from figure~\ref{bkg} that
we construct within the slow-rotation approximation. We also
show more ``realistic'' sequences based on conserving the 
total baryon number (see section~V). In addition, the figure shows that, just 
as for single-fluid stars, there will be a family
of supramassive stars that have no stable non-rotating counterpart \cite{cook}.

\begin{figure}[tbh]
\centerline{\epsfxsize=9cm \epsfbox{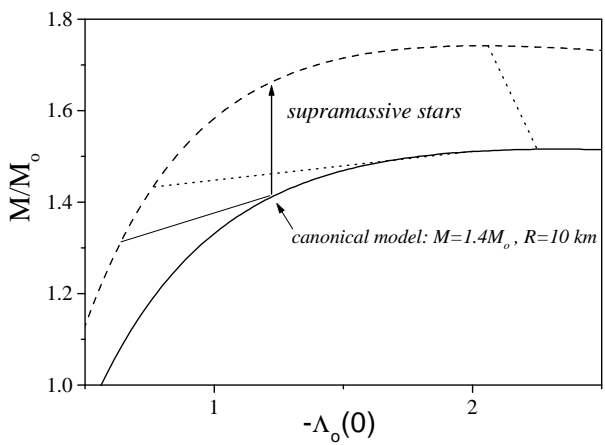}} 
\caption{We compare the mass of the non-rotating star (solid line) 
to its maximum value in the rotating case (which is 
reached when the star spins at the Kepler limit, dashed line)
for a sequence of stars parameterized by the central energy density 
$-\Lambda_{\rm o}(0)$.
The vertical arrow indicates rotating configurations with the 
same central energy density as the canonical model 
shown in figure~\ref{bkg}. The dotted line starting at the 
base of the arrow shows a more ``realistic'' sequence 
of rotating stars constructed by conserving the total 
baryon number. We also show a similar sequence corresponding to the 
maximum mass non-rotating star (upper dotted line).
Above this latter line, the various stars have no 
stable non-rotating counterpart, they are the so-called 
supramassive stars.}
\label{kepseq}
\end{figure}

Another illustration of the effects that rotation has on the 
inertial mass of the star is shown in figure~\ref{momseq}.
Here we show how the total mass for a star
rotating at  the Kepler limit depends on the relative
rotation between the neutrons and the protons. The particular
results are for our canonical stellar model from figure~\ref{bkg},
and the behaviour is easily understood if we compare the 
results for the mass to the Kepler frequency results in 
figure~\ref{kepom}.

\begin{figure}[tbh]
\centerline{\epsfxsize=9cm \epsfbox{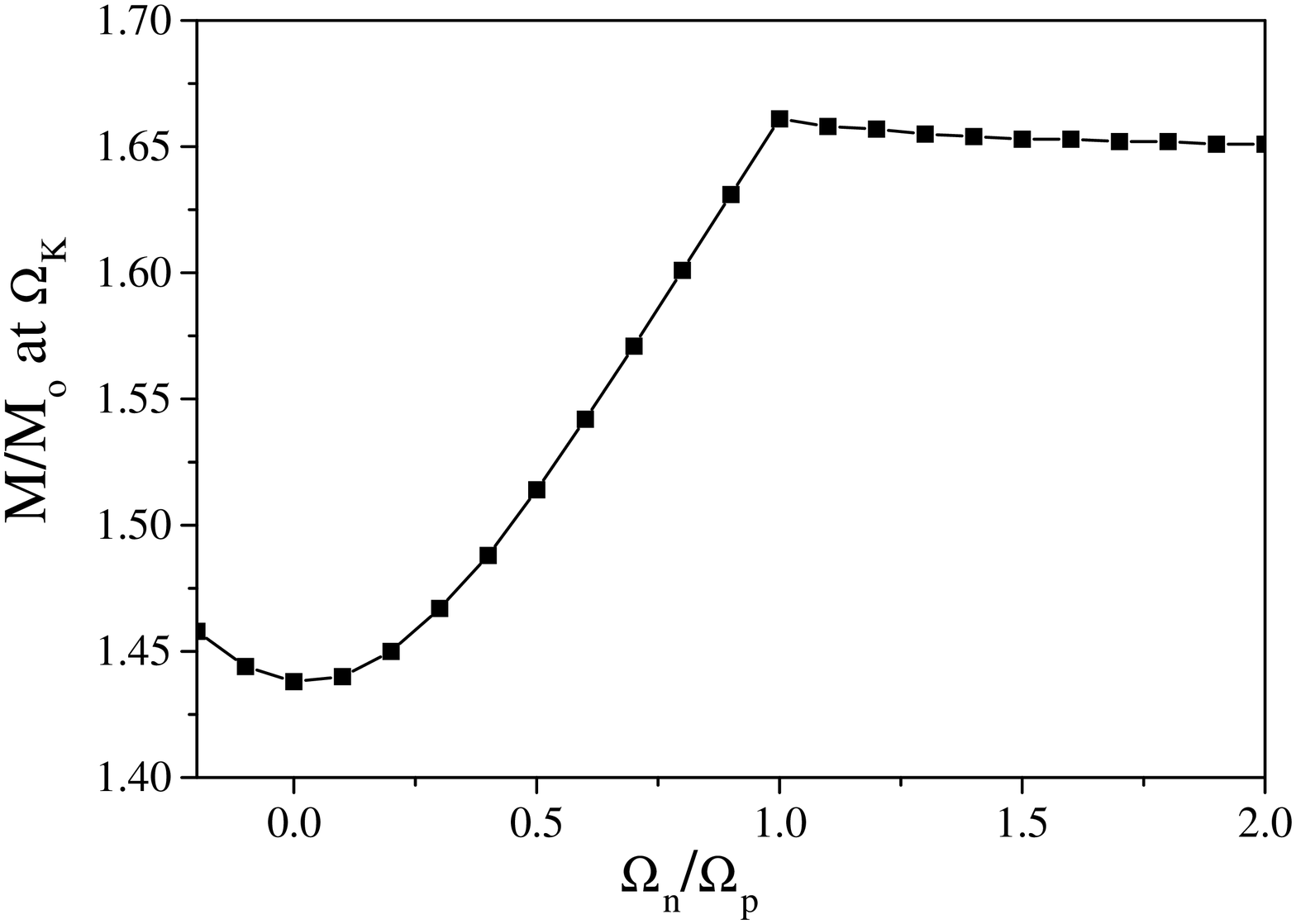}} 
\caption{ An illustration of how the total gravitational mass for a star
rotating at  the Kepler limit depends on the relative
rotation between the neutrons and the protons.
These particular results are for our canonical stellar model.}
\label{momseq}
\end{figure} 

The obtained solution for $h_0$ is 
illustrated in Figure~\ref{hk}. The variables 
$\xi_0$, $\eta_0$ and $\Phi_0$ that 
are solved for algebraically are shown in figures~\ref{xi}-\ref{fi}.
In these figures it is worth noticing that the rotational
correction to the neutron number density $\eta_0$ diverges at the 
surface. One can easily show that such a divergence will be present
whenever the index of the corresponding polytrope (here $\beta_n$)
is larger than 2. It should be pointed out, however, that this
does not affect the determination of any physical parameters
such as the total mass or the corresponding baryon number.
Finally, we note that the function $m_0$ increases 
monotonically from the centre to the surface. 
Hence, we do not illustrate it here.

\begin{figure}[tbh]
\centerline{\epsfxsize=9cm \epsfbox{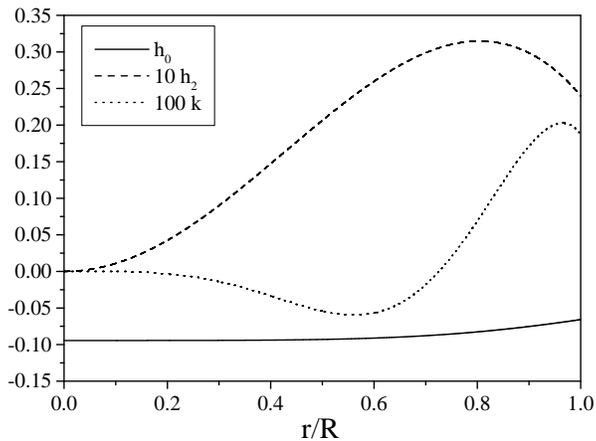}} 
\caption{The metric functions $h_0$, $h_2$ and $\tilde{k}$  for 
our model star and the case when the superfluid neutrons corotate 
with the protons ($\Omega_n=\Omega_p$). This particular data 
corresponds to the protons rotating at $\nu_p = 1$~kHz.}
\label{hk}
\end{figure}

\subsection{The $l = 2$ Results}

The final step in constructing a slowly rotating superfluid star
consists of solving the $l=2$ equations (\ref{coortrans2}), 
(\ref{flden2}), 
(\ref{222}), \ref{122}) and (\ref{112}).  These equations
determine $\xi_2$, $\eta_2$, $\Phi_2$, $h_2$, $v_2$ and
$k_2$. Equation (\ref{002}) 
contains no new information and we will not use it here. 

The strategy for solving these
equations is similar to that we adopted for the $l=0$ problem. 
In the $l=2$ case it is, however, useful to rewrite the equations
somewhat. We note that once a solution to the differential
equations (\ref{122}) and (\ref{112}) 
is obtained, all other variables follow algebraically. 
As Hartle points out \cite{H1}, it is convenient to introduce a new 
variable
$\tilde{k}=k_2+h_2$. This definition and some straightforward
manipulations lead to two coupled equations for $\tilde{k}$ and $h_2$.
These can be written
\begin{eqnarray} 
\tilde{k}^\prime &+& \nu^\prime h_2 = {r^3 \over e^\nu}
 \left( 2 + r \nu^\prime \right) \left\{ 
{1\over 12 e^\lambda}
\left( {d\tilde{L}_n \over dr} \right)^2 + {4\pi \over 3} \left[ 
(\Psi_0-\Lambda_0)\tilde{L}_n^2 + \chi_0 p_0(\tilde{L}_n+\tilde{L}_p)
(\Omega_n-\Omega_p)-A_0 n_0 p_0  (\Omega_n-\Omega_p)^2 \right]
 \right\} \ ,
\label{l21}\end{eqnarray} 
\begin{eqnarray} 
h_2^\prime &+& \left\{ \nu^\prime + {1 \over \nu^\prime}\left[ {2e^\lambda
-2 \over r^2} + 8\pi e^\lambda(\Lambda_0-\Psi_0) 
\right] \right\} h_2 + {4e^\lambda \over r^2 \nu^\prime} \tilde{k} = 
\nonumber 
\\ &=& {r^2 \over 12 e^\nu \nu^\prime} \left\{ \left( r^2 e^{-\lambda} 
(\nu^\prime)^2-2 
\right) \left( {d \tilde{L}_n \over dr} \right)^2 + 16\pi r^2 
(\nu^\prime)^2 (\Psi_0-\Lambda_0) \tilde{L}_n^2 
\right. \nonumber \\
&+& \left. 32\pi e^\lambda (\mu_0 n_0 \tilde{L}_n^2
+\chi_0 p_0 \tilde{L}_p^2)+
16\pi \chi_0 p_0 r^2 (\nu^\prime)^2 (\Omega_n-\Omega_p)(\tilde{L}_n+ 
\tilde{L}_p) -
16\pi n_0 p_0 A_0 \left( 2e^\lambda + r^2 (\nu^\prime)^2\right)
(\Omega_n-\Omega_p)^2
\right\} \ .
\label{l22}\end{eqnarray} 

By analyzing these equations for small values of $r$ we find that in order
to have a regular solution we must
require that $h_2 = c_1 r^2$ and $\tilde{k}=c_2 r^4$ near the centre
of the star. A series expansion
of (\ref{l22}) then reveals that the two 
constants $c_1$ and $c_2$ should be related according to
\begin{eqnarray}
&& c_2+2\pi \left( \Psi_0 - {1 \over 3} \Lambda_0 \right) c_1 = \nonumber \\
&& {2 \pi \over 3 e^{\nu}} \left[  2(\Psi_0-\Lambda_0)\tilde{L}_n^2 - n_0 p_0 A_0
(\Omega_n-\Omega_p)^2 - 2\chi_0 p_0 (\Omega_n-\Omega_p)(\tilde{L}_n+\tilde{L}_p) - (\mu_0 n_0 \tilde{L}_n^2 + \chi_0 p_0 \tilde{L}_p^2) \right] \ .
\end{eqnarray}
Furthermore, since we know that $\xi_2\to 0$ as $r\to 0$ we can deduce 
from (\ref{coortrans2}) and {\ref{flden2}) that we must have
$\eta_2 = \Phi_2 =0$ at the centre of the star. 

Schematically, we solve the $l=2$ equations as follows: 
First we find a solution to (\ref{l21}) and (\ref{l22}) that matches
smoothly onto the vacuum solution at the surface of the star 
(cf. Section~IIIC). 
We do this by finding a particular solution to the full problem
and adding to it a multiple of the solution to the corresponding
homogeneous problem, where the right hand sides of (\ref{l21}) and 
(\ref{l22}) are set equal to zero and the two constants are 
related by
\beq
c_2+2\pi \left( \Psi_0 - {1 \over 3} \Lambda_0 \right) c_1 = 0 \ .
\eeq
The desired solution for $[\tilde{k},h_2]$ is then  the linear combination
that gives the vacuum relation between these two variables
at the surface. Having determined the appropriate linear combination
of the interior functions
(as well as the constant $A$ in the equations in Section~IIIC)
we can readily reconstruct the full solution, and in addition 
solve (\ref{coortrans2}) and ({\ref{flden2}) for 
$\xi_2$, $\eta_2$ and $\Phi_2$ after each integration step.
This then completes the construction of a slowly
rotating superfluid stellar model. Typical results for our 
model star are shown in figures~\ref{hk}-\ref{fi}. 

Once we have solved all the $l=2$ equations we can work out the
rotationally induced change in the shape of the star, i.e.
the centrifugal flattening. This effect 
is conveniently expressed in terms of the ratio between the polar and 
equatorial radii, $R_p$ and $R_e$ respectively. Within the slow-rotation 
approximation this leads to
\beq
{R_p \over R_e} \approx 1 + {3\xi_2(R) \over 2R}
\eeq
and for our model star (with $\eta_0(0)=0$ and $\Omega_n=\Omega_p$)
we get
\beq
{R_p \over R_e} \approx 1 - 0.113 \left( {\nu_p \over 1~\mbox{kHz}} 
\right)^2  \ .
\eeq
Thus, we find $R_p/R_e\approx 0.7$ for a star spinning at the 
mass-shedding limit.

\begin{figure}[tbh]
\centerline{\epsfxsize=9cm \epsfbox{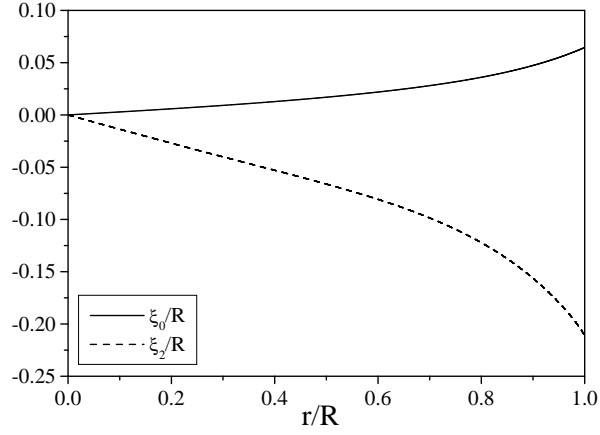}} 
\caption{ The functions $\xi_0$ and $\xi_2$ for 
our model star and the case when the superfluid neutrons corotate 
with the protons ($\Omega_n=\Omega_p$).This particular data 
corresponds to the protons rotating at $\nu_p = 1$~kHz. }
\label{xi}
\end{figure}

\begin{figure}[tbh]
\centerline{\epsfxsize=9cm \epsfbox{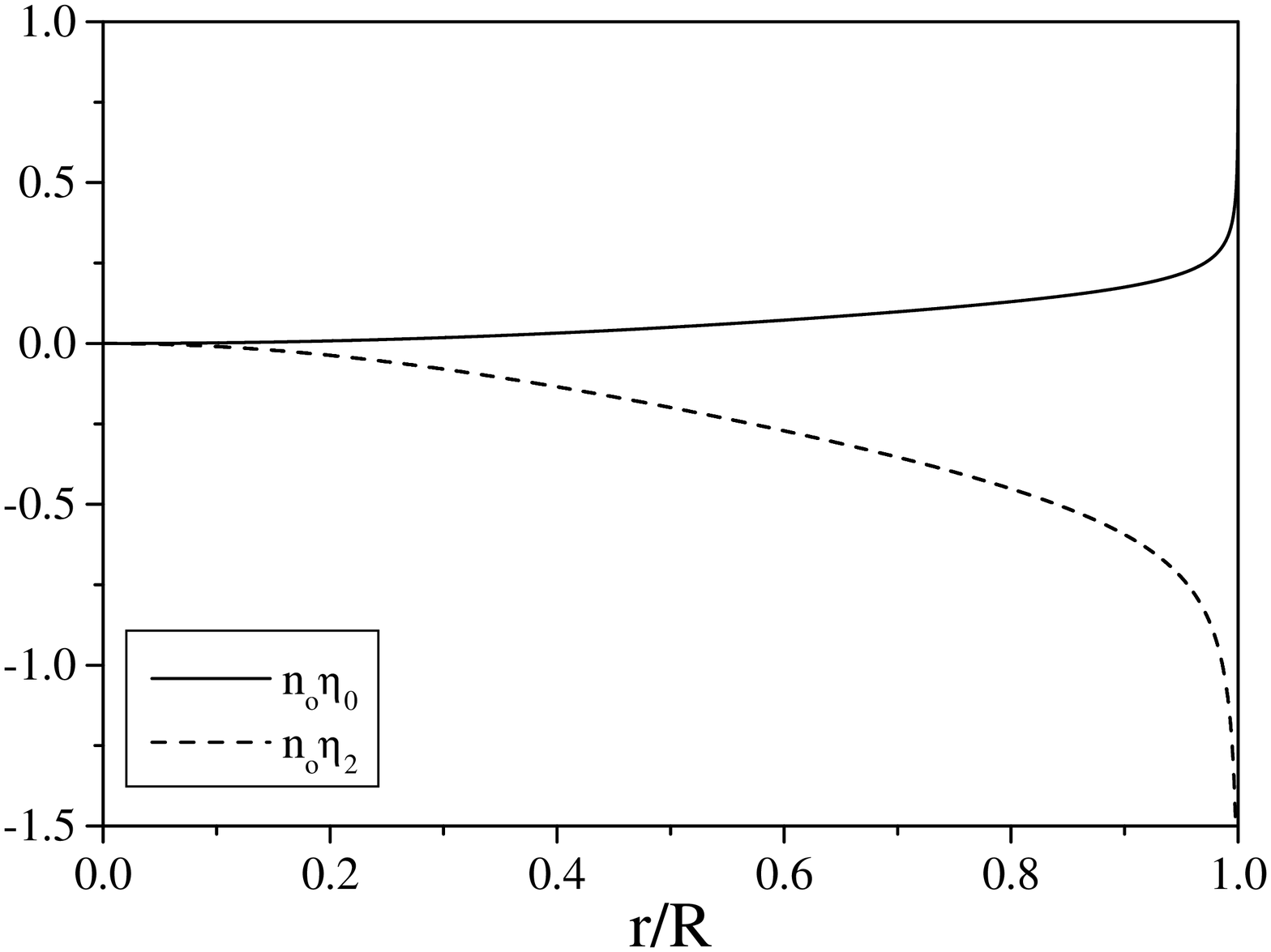}} 
\caption{ The rotationally induced changes in the neutron number 
density, $\eta_0$ and $\eta_2$, for 
our model star and the case when the superfluid neutrons corotate 
with the protons ($\Omega_n=\Omega_p$).This particular data 
corresponds to the protons rotating at $\nu_p = 1$~kHz. }
\label{eta}
\end{figure}

\begin{figure}[tbh]
\centerline{\epsfxsize=9cm \epsfbox{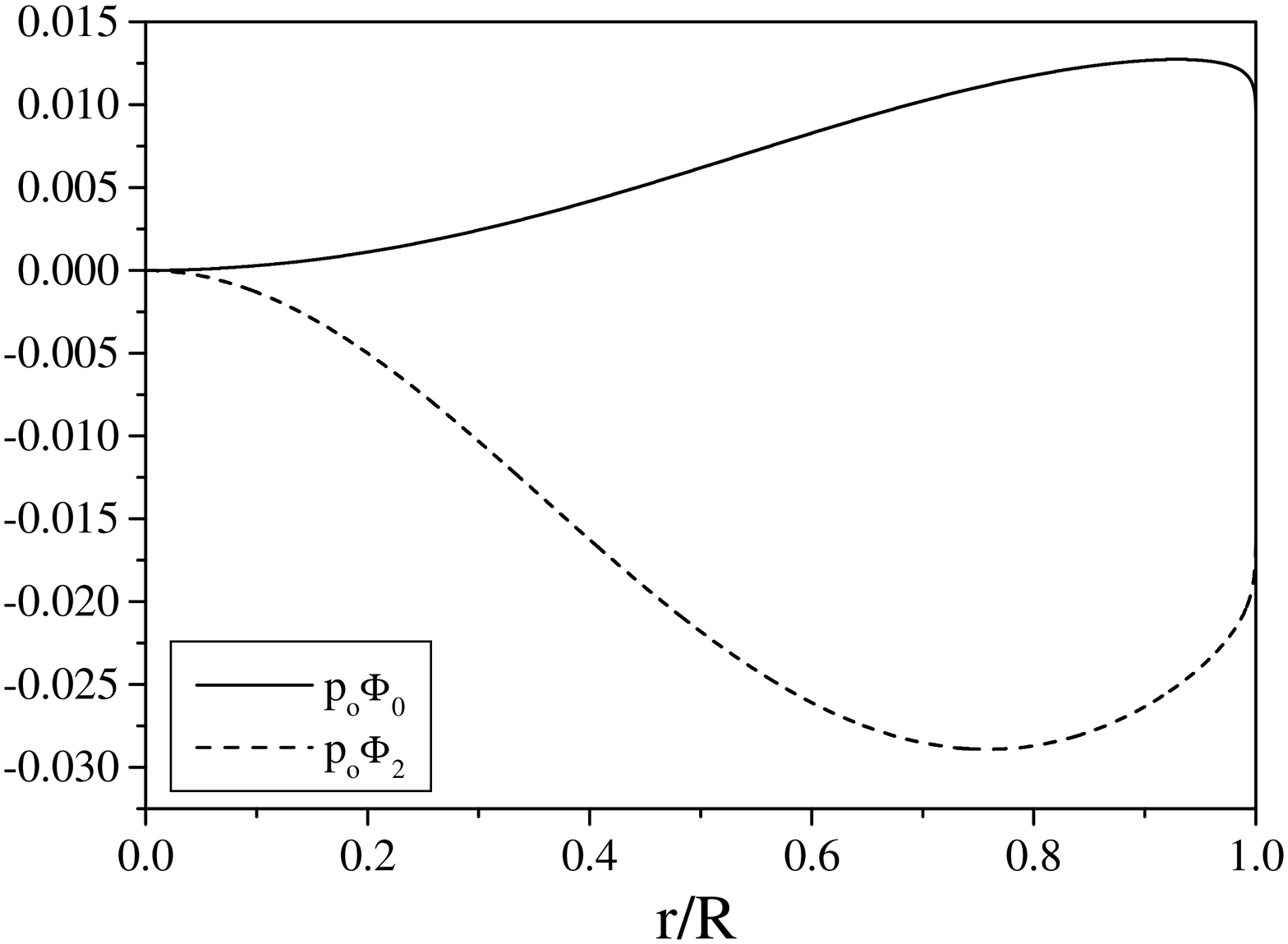}} 
\caption{ The rotationally induced change in the proton number 
density, $\Phi_0$ and $\Phi_2$, for 
our model star and the case when the superfluid neutrons corotate 
with the protons ($\Omega_n=\Omega_p$). This particular data 
corresponds to the protons rotating at $\nu_p = 1$~kHz.}
\label{fi}
\end{figure}

Once we have solved the $l=2$ equations and determined the 
value of $A$ we can calculate the quadrupole moment from (\ref{qpole}).
In doing this it is useful to construct
two dimensionless quantities
\beq
q = \left( {c^2 \over GM} \right) {Q \over M} \quad , \mbox{ and }
 \quad \chi = {c \over G} {J \over M^2} \ .
\eeq
Once we do this, we can compare our results to those of 
Laarrakers and Poisson \cite{LP} who study the quadrupole moment
of rapidly rotating single fluid stars for realistic 
equations of state. In figure~\ref{quadru} we compare 
our results to theirs by showing the dependence of
the ratio $q/\chi^2$ on the mass of the star.
Clearly, the present results for the quadrupole moment 
are reasonable. It should also be noted that 
one would have  $q/\chi^2= -1 $ for a Kerr black hole. Thus
figure~\ref{quadru} emphasizes the well-know fact that 
rotating neutron stars have a multipole structure that is radically 
different from that of the Kerr spacetime.

\begin{figure}[tbh]
\centerline{\epsfxsize=9cm \epsfbox{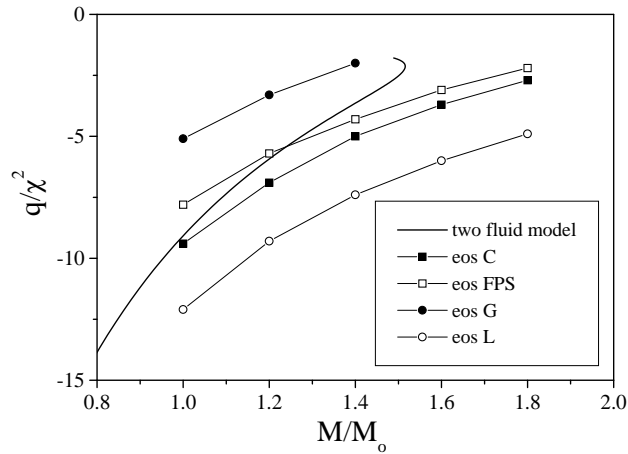}} 
\caption{Our results for the quadrupole moment (in terms of the 
ratio $q/\chi^2$, see the main text for definitions) as a function
of the stellar mass are compared to numerical results obtained for 
rapidly rotating relativistic stars using realistic 
supranuclear equations of state.}
\label{quadru}
\end{figure}

\section{Sequences of stars}

Given the machinery we have developed in the previous sections we
are set to ask physically relevant questions. 
Ideally,
we would like to be able to construct a sequence of rotating models
that can meaningfully be said to represent the same non-rotating star.
Given that we can allow the neutrons and the protons to rotate
differently, and also  adjust the relative central densities,
the construction of such a sequence of models
is non-trivial. However, a natural starting point may be to 
first determine the maximum allowed rate of rotation. 
Then we will discuss 
conservation of the total integrated neutron and proton numbers
along a sequence of rotating stars. To impose this requirement 
corresponds to 
determining the total baryon numbers for the neutrons and the protons
and then finding models such that these quantities are held
constant as the spin-rate increases.
Finally, we will turn to issues regarding chemical equilibrium.

\subsection{The Kepler limit}

The maximum spin-rate of a stationary configuration corresponds to the
so-called Kepler frequency. It is determined as the rotation rate where 
mass-shedding sets in at the equator. The corresponding frequency
can be calculated by requiring that the rotation rate of the fluid be 
equal to the frequency of a particle orbiting the star on an equatorial 
geodesic.
The determination of the Kepler frequency for our slowly 
rotating stars is relatively straightforward. After all, the 
main part of the calculation involves the rotation 
frequency of an orbiting particle, and as far as the exterior
vacuum is concerned, the fact that we are dealing with 
the rotation of two coupled fluids is irrelevant. The only subtlety
of the two-fluid problem involves the fact that we can allow the
protons and neutrons to rotate at different rates. Clearly, the
Kepler limit will be determined by whichever fluid is rotating the fastest
at the surface. In other words, we simply take $\Omega_K = \mbox{max }(
\Omega_n,\Omega_p)$ in the following. It should, of course, be noted
that the situation is simpler for more realistic neutron star models. 
Then the Kepler frequency will be determined from the rotation rate of the
crust. 

For a rapidly rotating star the Kepler frequency $\Omega_K$ can be 
determined from 
(cf. Friedman, Ipser and Parker\cite{fip})
\begin{equation}\Omega_K = {N v \over \sqrt{K}} + \omega 
\end{equation} 
where the metric variables are defined by (\ref{finmetric})
and
\begin{equation}
v = {K^{3/2} \omega^\prime \over N K^\prime} + \sqrt{ {2K N^\prime \over
N K^\prime} + \left({ K^{3/2}\omega^\prime \over N K^\prime} \right)^2 }  
\end{equation}
is the orbital velocity according to a zero-angular momentum observer at the 
equator (where all quantities should be evaluated). Here primes denote
derivatives with respect to $\tilde{r}$.
In order to be consistent, we expand this equation keeping only 
terms up to (and including) order $\Omega^2$. Then we get
\begin{equation}
\Omega_K \approx { e^{\nu/2} \over \tilde{r}} \sqrt{ {\tilde{r}
\nu^\prime \over 2}} 
+ \omega + {\tilde{r} \omega^\prime \over 2} +  
e^{\nu/2}  \sqrt{ {\nu^\prime \over 2\tilde{r}}} \left[ h-k+{h^\prime 
\over \nu^\prime} - {\tilde{r}k^\prime \over 2} +
{\tilde{r}^3(\omega^\prime)^2 \over 4 \nu^\prime e^\nu} \right]
+O(\Omega^3) \ .
\label{kep2}\end{equation}
The next step corresponds to using the vacuum expressions for the 
various slow-rotation quantities, and also doing the coordinate
transformation $\tilde{r}\to r$. Luckily, the latter affects 
only the first term in (\ref{kep2}). After some algebra we arrive at 
our final expression
\begin{equation}
\Omega_K = \sqrt{M \over R^3} - {\hat{J}\Omega_\p \over R^3} + \sqrt{M \over R}
\left\{ {\delta \hat{M} \over 2M} + { (R+3M)(3R-2M) \over 4R^4M^2} \hat{J}^2 -
{3\over 4} {2\hat{\xi}_0+\hat{\xi}_2 \over R^2} + 
\alpha \hat{A} \right\} \Omega_\p^2
\label{kep3}\end{equation}
where we have made the scaling with $\Omega_p$ explicit by introducing
$J=\hat{J} \Omega_p$ etc.
We have also defined
\begin{equation}
\alpha = {3(R^3-2M^3) \over 4M^3} \log \left( 1-{2M \over R} \right)
+ { 3R^4-3R^3M-2R^2M^2-8RM^3+6M^4 \over 2RM^2(R-2M)} \ .
\end{equation}
Here it should be noted that, at the $\Omega^2$ level the Kepler
limit depends explicitly on the rotationally induced change in mass, the 
centrifugal flattening as well as the change in quadrupole moment
(through $A$).

In order to determine the Kepler limit for our rotating 
superfluid models we simply have to solve the above quadratic for 
$\Omega_p$. This then tells us the maximum allowed spin-rate 
for the protons at any given central energy density and relative
(constant) rotation rate $\Omega_\n/\Omega_\p$. Typical results 
of this exercise are shown in figure~\ref{kepom}.
In this figure we show how the Kepler frequency is affected by changes
in the relative rotation rate $\Omega_\n/\Omega_\p$. 
The results can be understood in the following way: For 
$\Omega_\n/\Omega_\p=1$ the two fluids rotate together so we have a 
fairly standard result. For the model star from figure~\ref{bkg}
we find that the Kepler frequency is 
$\Omega_K=\Omega_\p=\Omega_\n\approx 1.02\times 10^4$~rad/s or 
$\nu_K \approx 1630$~Hz. When we compare this to the 
relation (\ref{empire}) we see that it is not an unreasonable
result, although it is roughly 10\% higher than the empirical estimate
for the relevant mass and radius. Figure~\ref{kepom} also provides
information regarding the effects of differential rotation. 
First of all it is clear that when $\Omega_\n>\Omega_\p$ the Kepler
rate is determined by the neutrons. Given that the neutrons 
make up more than
90\% of the material in the star it is not too surprising that 
the result for $\Omega_K$ changes very little as we increase 
$\Omega_n$ beyond $\Omega_p$. Basically, the star rotates like a single
fluid star with a dynamically small proton component. 
The opposite case, when $\Omega_\n < \Omega_\p$, is more interesting. 
Then we see that the maximum allowed spin rate for the 
protons increases as $\Omega_\n$ decreases. We can deduce that this
is reasonable from (\ref{kep3}): As we decrease the rotation rate
of the neutrons the various $\Omega_\p^2$ terms all decrease (again,
because the neutrons make up most of the star and so have a dominant
effect). Thus it follows naturally that the orbital frequency
of a particle at the equator will approach the non-rotating star
case. In other words, it increases which leads to the maximum 
allowed rotation of the protons increasing as well.

\begin{figure}[tbh]
\centerline{\epsfxsize=9cm \epsfbox{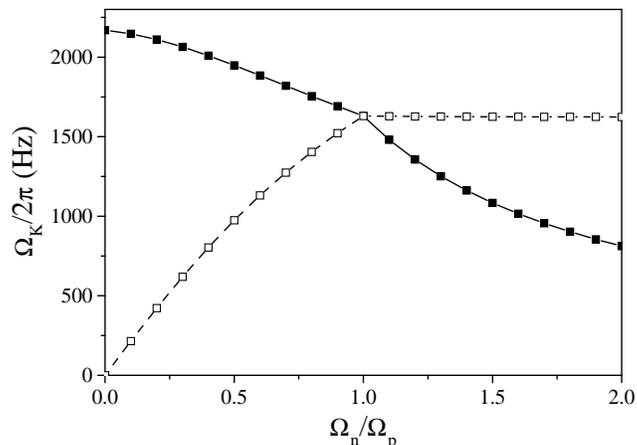}} 
\caption{The Kepler limit $\Omega_K$ is shown as a 
function of the relative rotation rate $\Omega_n/\Omega_p$
for our model star. The filled squares show the maximum 
allowed rotation rate for the protons ($\Omega_p$), while 
the open squares show the corresponding neutron spin rate
($\Omega_n$). The Kepler frequency simply corresponds to the 
largest of the two.}
\label{kepom}
\end{figure}

\subsection{The baryon numbers and angular momenta}

We want to determine  the conserved 
neutron number $N_{\n}$, the conserved proton number $N_{\p}$, and then 
define the total baryon mass $M_B$ as
\beq
    M_B = m_{\n} N_{\n} + m_{\p} N_{\p} \ ,
\eeq
where $m_{\n}$ is the neutron mass and $m_{\p}$ is the proton mass.
One of the assumptions behind the general relativistic superfluid 
formalism implemented here is that the neutrons and the protons are 
independently conserved. 
Thus the formulas for the conserved particle numbers can be obtained 
by integrating the 
conservation rules (\ref{coneq}) over a bounded region of spacetime, 
and then using the 
generalized Stokes Theorem to reduce the spacetime volume integral to 
three-dimensional integration over the boundary.  The bounded 
spacetime region is the four-dimensional volume contained 
between the two spacelike slices given by $t = t_1$ and $t = t_2$ where 
$t_1$ and $t_2$ are constants.  The timelike boundary of this volume is 
the union of each sphere at infinity that exists on each $t =$~constant 
spacelike slice.  Because the matter is assumed to have compact support, 
there will be no contributions to the baryon mass formula from the 
timelike boundary.  All that remains are two integrals, 
one on the $t = t_1$ and another on the $t = t_2$ hypersurface, that 
add to zero.  The $t = t_2$ integral, say, then yields the conserved 
particle number. 

The conserved neutron particle number is thus found to be 
\beq
    N_{\n} = - \int_{t = const} \n^{\mu} \eta_{\mu} \sqrt{\gamma} 
             {\rm d}^3x \ ,
\eeq
where $\gamma = {\rm det} \gamma_{i j}$ and $\gamma_{i j}$ is the 
intrinsic metric of the $t =$~constant hypersurface, i.e. the metric is 
assumed to have a $3+1$ decomposition of the form 
\beq
    {\rm d}s^2 = - \left(N^2 - \gamma_{ij} N^i N^j\right) {\rm d}t^2 
                 + 2 \gamma_{ij} N^j {\rm d}x^i {\rm d}t + \gamma_{ij} 
                 {\rm d}x^i {\rm d}x^j \ ,
\eeq
and the (future-directed) normal to the $t =$~constant hypersurfaces, 
to be denoted $\eta^{\mu}$, is given by $\eta_{\mu} = (- N,0,0,0)$.  
The formula for $N_{\p}$ is identical except that $\n^{\mu}$ is 
replaced with $\p^{\mu}$.  One important point to keep in mind is that 
the surface of the star (which will be non-spherical in general) is 
given by $r =$~constant, cf. Figure~\ref{isos}.
In terms of the metric and particle number density currents developed 
for the slow rotation formalism it is found that 
\beq
   N_{\n} = 4 \pi \int_0^R {\rm d} r r^2 e^{\lambda/2} 
              \n_{\rm o} \left(1 + \eta_0 + v_0 + \left[{
              \lambda^{\prime} \over 2} + {2 \over r}\right] \xi_0 
              + {2 r^2 \over 3 e^{\nu}} \tilde{L}^2_{\n}
              \right) \ ,
\eeq
\beq
   N_{\p} = 4 \pi \int_0^R {\rm d} r r^2 e^{\lambda/2} 
              \p_{\rm o} \left(1 + \Phi_0 + v_0 + \left[{
              \lambda^{\prime} \over 2} + {2 \over r}\right] \xi_0 
              + {2 r^2 \over 3 e^{\nu}} \tilde{L}^2_{\p}
              \right)   \ .
\eeq
As an illustration we note that in the case of the model
in Figure~\ref{bkg}, and $\Omega_\n=\Omega_\p$, the two baryon masses
are  
\begin{eqnarray*}
M_{Bn}/M_\odot &\approx& 1.46 +  0.13\left({\nu_\p \over 1~\mbox{kHz}} 
\right)^2 \\ 
M_{Bp}/M_\odot &\approx& 0.11 +  0.027\left({\nu_\p \over 1~\mbox{kHz}} 
\right)^2 \ .
\end{eqnarray*}

In the standard single fluid case one would typically identify a 
distinct sequence of rotating models as representing ``the same''
star by requiring that the total baryon rest mass be conserved.
In our present problem, the fact that we are dealing with 
two distinct fluids provides additional complications. 
Provided that the star spins up (or down) rapidly enough 
compared to the nuclear reactions that convert neutrons
to protons (and vice versa) it would be reasonable to conserve
the total integrated neutron and proton neutron numbers
individually. However, the spin-evolution of a neutron
star typically proceeds slowly. Furthermore, as discussed by 
Reisenegger \cite{reis}, one would expect a certain amount of 
conversion between the two fluids as the star spins down simply 
because of the change in shape of the constant density 
surfaces. In other words, in the two fluid case it is not
obvious how to construct an ``astrophysical sequence'' of 
rotating stars based on the two baryon masses. 

Anyway, 
let us suppose that we want to demand that the individual
integrated baryon numbers are conserved as we spin the star up.  
From the above results we see that this corresponds to
\beq
\delta N_n = \int_0^R {\rm d} r r^2 e^{\lambda/2} 
              \n_{\rm o} \left(\eta_0 + v_0 + \left[{
              \lambda^{\prime} \over 2} + {2 \over r}\right] \xi_0 
              + {2 r^2 \over 3 e^{\nu}} \tilde{L}^2_{\n}
              \right) = 0 \ , 
\label{baryn}\eeq
and
\beq
 \delta N_p = \int_0^R {\rm d} r r^2 e^{\lambda/2} 
              \p_{\rm o} \left(\Phi_0 + v_0 + \left[{
              \lambda^{\prime} \over 2} + {2 \over r}\right] \xi_0 
              + {2 r^2 \over 3 e^{\nu}} \tilde{L}^2_{\p}
              \right) = 0   \ .
\label{baryp}\eeq
Let us consider the (by now familiar) stellar model from 
figure~\ref{bkg}.  According to the result above we then have 
$$
M_B=M_{Bn}+M_{Bp} = (1.46+0.11) M_\odot = 1.57 M_\odot \ ,
$$
in the rotating case. 
First of all we find that the two baryon masses are not
individually conserved for the star that has the same
total $M_B$ at the Kepler limit. At least not if we take
$\eta_0(0)= 0 $ and $\Omega_n/\Omega_p=1$. For this case
we find
$$
M_{Bn} =  1.48 M_\odot\ , \qquad \mbox{ and } \qquad M_{Bp} = 
0.09M_\odot \ .
$$
However, we have the freedom to choose our parameters differently
and we find that we can conserve both baryon masses 
by adjusting either  $\eta_0(0)$ or $\Omega_n/\Omega_p$
(or, indeed, both of them). For example, we find that 
both $M_{Bn}$ and $M_{Bp}$ are the same as in the 
non-rotating star if we take
\begin{eqnarray*}
\eta_0(0) &=& -4\times 10^{-2} \left({\nu_\p \over 1~\mbox{kHz}} 
\right)^2 
\quad \mbox{ and } \quad \Omega_n/\Omega_p = 1 \ , \\
\eta_0(0) &=& 0 \quad \mbox{ and } \quad \Omega_n/\Omega_p = 0.75 \ .
\end{eqnarray*}
This is an interesting illustration of the subtleties involved
in constructing rotating superfluid neutron star configurations.

Furthermore, the fact that the slow-rotation 
calculation requires, once
a non-rotating star is given, the specification of 
both $\Omega_n/\Omega_p$ and  $\eta_0(0)$ leads us to 
ask an intriguing question: We have two seemingly free 
parameters and two equations, (\ref{baryn}) and (\ref{baryp}).
Is it possible to specify these constants in such a way that
the two  baryon numbers are conserved even though we 
keep the central density of the star fixed? 
In the standard one-fluid case this would certainly 
not be possibly. As the spin of the star
increases the centrifugal flattening will necessarily
lead to decrease in the central density (for a star with a 
fixed number of baryons). Is it possible that 
 two-fluid models offer a somewhat counter-intuitive
alternative? That is,  is it possible to carefully redistribute
the neutrons and the protons (in such a way that the central 
energy density remains unchanged), and spin the star up
without causing the familiar increase in baryon mass?
For our particular model equation of state we find a
negative answer to this question, cf. Figure~\ref{bary}.
However, it is an intriguing possibility and we cannot
at this point argue why it could not happen for some
other, perhaps more realistic, equation of state.

\begin{figure}[tbh]
\centerline{\epsfxsize=9cm \epsfbox{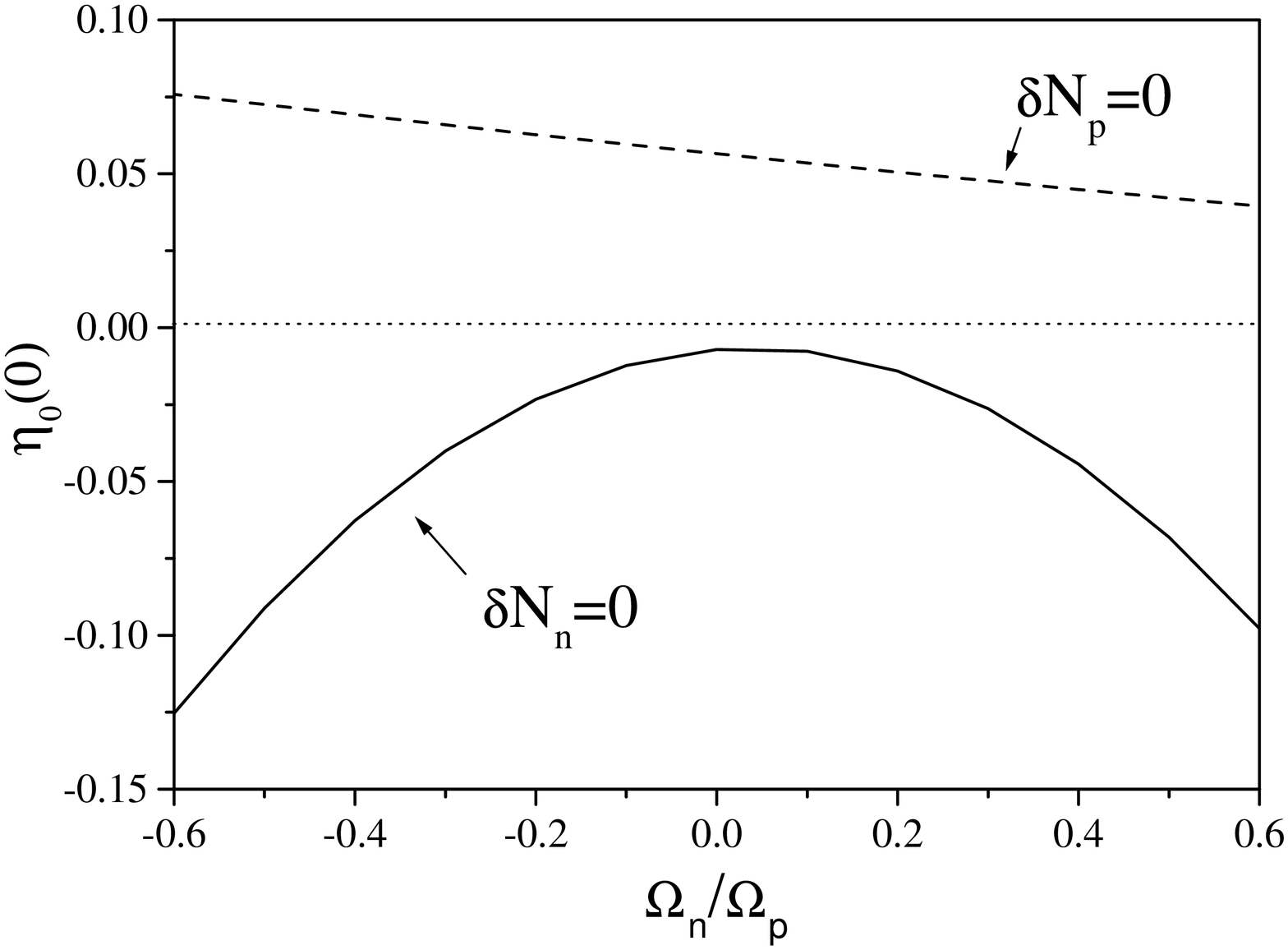}} 
\caption{Curves corresponding to conservation of the total 
number of neutrons ($\delta N_n=0$) and the total number 
of protons ($\delta N_p=0$) are shown as functions of the 
two free parameters $\eta_0(0)$ and $\Omega_n/\Omega_p$.
(The data corresponds to the case $\nu_p=$1~KHz.)
A point of intersection would correspond to a rotating model
which conserves both the individual baryon masses from 
the non-rotating star.
As is clear from the data, such models cannot be constructed 
for our model equation of state (which agrees with the standard
result from the one-fluid case).
This does not, however, rule out this possibility for other
equations of state. }
\label{bary}
\end{figure}

Using the above set-up for the bounded spacetime region, we can also 
make meaningful definitions for the individual neutron and proton total 
angular momenta, to be denoted $J_\n$ and $J_\p$ respectively.  Langlois 
et al \cite{LSC98} demonstrate that such an unambiguous separation can be 
made if the hypersurface over which the integrals are being performed is 
invariant under the axisymmetry action.  In the present context this 
means that the scalar product of the spacelike Killing vector 
$\phi^{\mu}$ with the normal to our hypersurface $\eta^{\mu}$ must be 
zero.  Indeed, it is trivial to verify that we have $\phi^{\nu} 
\eta_{\nu} = 0$.  Now, according to Langlois et al, the two angular 
momenta are given by
\beq
    J_\n = - \int_{t = const} \left(\phi^{\nu} \mu_{\nu}\right) 
           \n^{\mu} \eta_{\mu} \sqrt{\gamma} {\rm d}^3x 
           \quad , \quad 
    J_p = - \int_{t = const} \left(\phi^{\nu} \chi_{\nu}\right)
          \p^{\mu} \eta_{\mu} \sqrt{\gamma} {\rm d}^3x \ .
\eeq
In terms of the slow-rotation approximation, we find for the neutron 
total angular momentum 
\beq
J_{\n} = - {8 \pi \over 3} \int_0^R {\rm d}r r^4 e^{(\lambda - \nu)/2}
         \left[\mu_{\rm o} \n_{\rm o} \tilde{L}_{\n} + \A_{\rm o}
         n_{\rm o} p_{\rm o} \left(\Omega_{\n} - \Omega_{\p}\right)\right]
\eeq
and the very similar form
\beq
J_{\p} = - {8 \pi \over 3} \int_0^R {\rm d}r r^4 e^{(\lambda - \nu)/2}
         \left[\chi_{\rm o} p_{\rm o} \tilde{L}_{\p} + \A_{\rm o}
         n_{\rm o} p_{\rm o} \left(\Omega_{\p} - \Omega_{\n}\right)\right]
\eeq
for the proton total angular momentum.  It is easy to verify that 
\beq
    J = J_{\n} + J_{\p} \ ,
\eeq
where $J$ is the total angular momentum found earlier by matching to 
the exterior vacuum solution of Hartle.

\subsection{Chemical Equilibrium}

So far we have not imposed chemical equilibrium in our rotating 
stars (although it was assumed for the non-rotating background
model). This is quite reasonable since one can easily think of situations
where the two fluids are temporarly rotation ``out of sync'' 
following for example a glitch. As we argued in the Introduction
it will then likely take the various relevant mechanisms
many hundreds of dynamical timescales to again lock the 
two components together. And it may take much longer for chemical
equilibrium to be restored given that this happens on the
characteristic timescale of the various nuclear reactions.
Anyway, before we conclude this paper it  
is obviously meaningful to address the issue
of chemical equilibrium. 
Intuitively, one might expect that one can only have 
equilibrium when the neutrons and the protons corotate. 
The argument for this is simple: If the two species rotate
at different rates we must work in one of the two frames 
(the nuclear reactions that re-establish equilibrium 
depend on the local physics), and the result will depend on which
frame we choose. In the frame rotating with the neutrons the 
proton chemical potential will have an additional kinematic piece. 
This is  also true for the neutron chemical potential in the frame 
that corotates with the protons. In other words, it would seem
possible to have equilibrium only if the two fluids corotate.

This is, of course, just a hand-waving argument and we need to 
support it mathematically.  Let us first return to a situation 
where $\Omega_\n$ and $\Omega_p$ are not assumed to be constant.
Langlois et al \cite{LSC98} have shown that chemical 
equilibrium is imposed on the system via the condition
\beq
    \p^{\nu} \left(\mu_{\nu} - \chi_{\nu}\right) = 0 \ .
\eeq
Using the results from Section~II we immediately find that this 
corresponds to
\beq
t^\nu \mu_\nu + \Omega_p \phi^\nu \mu_\nu = 
		t^\nu \chi_\nu + \Omega_p \phi^\nu \chi_\nu \ .
\eeq
Let us now assume, again, that $\Omega_p$ is constant.
As we have already argued in Section~IV, this is likely to 
be the case in most astrophysical neutron stars. 
The Euler equation 
for the protons (\ref{eueqn}) then implies 
that the right-hand-side of the 
above condition is a constant. Thus we have
\beq
t^\nu \chi_\nu + \Omega_p \phi^\nu \chi_\nu = \mbox{ constant} \ .
\eeq
Obviously, this also implies that 
\beq
t^\nu \mu_\nu + \Omega_p \phi^\nu \mu_\nu = \mbox{ constant} \ .
\eeq
We solve for $t^\nu \mu_\nu$ and then put the result into the Euler 
equation for the neutrons. This way we obtain
\beq
(\Omega_n - \Omega_p) \partial_\mu (\phi^\nu \mu_\nu) = 0 \ . 
\eeq

From this we can conclude the following:
If one demands chemical equilibrium, 
and also assumes that $\Omega_p$ is constant (that the protons
rotate rigidly), then one of the following two conditions must be true,
\beq
\Omega_n - \Omega_p = 0 \ ,
\eeq
or
\beq
\phi^\nu \mu_\nu = \mbox{ constant} \ .
\eeq
The first of these is the result that we anticipated, that 
chemical equilibrium would lead to the two fluids co-rotating. 
The second condition is a bit more puzzling, and we need to 
ask whether one can have a physical configuration that  
satisfies it.  

Taking the  slow-rotation form for $\mu_\nu$ 
the second condition becomes
\beq
 - r^2 {\rm sin}^2\theta e^{- \nu/2} [{\cal B}_{\rm o} n_{\rm o} 
           (\omega - \Omega_n) +
           {\cal A}_{\rm o} p_{\rm o} (\omega - \Omega_p)] = 
           \mbox{ constant} \ .
\eeq
From this it is clear that unless the right-hand-side vanishes then 
$\omega$ and  $\Omega_n$ are both badly
behaved at the origin (as well as at the poles).
Thus we can only have a well-behaved solution if
\beq
\Omega_n = [({\cal B}_{\rm o} n_{\rm o} + A_{\rm o} p_{\rm o}) 
           /({\cal B}_{\rm o} n_{\rm o})] \omega -
	   [A_{\rm o} p_{\rm o}/{\cal B}_{\rm o} n_{\rm o}] \Omega_p \ .
\label{newcon}\eeq
In the special case with no entrainment we have 
${\cal A}_{\rm o}=0$, and our condition
implies
\beq
\Omega_n = \omega \ .
\label{newchem}
\eeq
Hence, we see that the second 
condition for  chemical equilibrium condition 
 links the rotation rate of the superfluid
neutrons to the frame-dragging. This means that, in order 
to satisfy this condition, the neutrons must rotate 
differentially. At first this possible solution may seem
peculiar, but it may in fact make sense if we consider 
the physical meaning of the frame-dragging. The condition 
(\ref{newchem}) simply says that the neutrons are not
rotating with respect to a local zero-angular momentum
observer. They are simply dragged along by the 
rotation of the protons.   

From this discussion we conclude that if the two fluids 
both rotate uniformly
and $\Omega_n\neq \Omega_p$ then one cannot impose chemical 
equilibrium.  When chemical equilibrium is 
imposed, the two fluids must either corotate, or $\Omega_n$ 
must have differential rotation (and in fact be
equal to the frame-dragging $\omega$ when there is no entrainment).

We note that one could, in principle, explore 
the latter possibility further. We can combine 
the condition for equilibrium with the frame-dragging equation 
(\ref{frmdrg}) in an interesting way. First we re-express (\ref{frmdrg})
in terms of $\tilde{L}_p$. Then we get 
\beq
    {1 \over r^4} \left(r^4 e^{- (\lambda + \nu)/2} 
      \tilde{L}_{\p}^{\prime}\right)^{\prime} - 16 \pi e^{(\lambda - 
      \nu)/2} \left(\Psi_{\rm o} - \Lambda_{\rm o}\right) \tilde{L}_{\p} 
      = 16 \pi e^{(\lambda - \nu)/2} \mu_o n_o 
      \left(\Omega_{p} - \Omega_{n}\right) \ .
\eeq
By making use of (\ref{newcon}) we readily rewrite this as
\beq
    {1 \over r^4} \left(r^4 e^{- (\lambda + \nu)/2} 
      \tilde{L}_{\p}^{\prime}\right)^{\prime} - 16 \pi e^{(\lambda - 
      \nu)/2} \left[ \Psi_{\rm o} - \Lambda_{\rm o} + 
{\mu_{\rm o} \over {\cal B}_{\rm o}} \left( {\cal B}_{\rm o} n_{\rm o} + 
{\cal A}_{\rm o} p_{\rm o} \right)\right]\tilde{L}_{\p} 
      = 0 \ .
\label{newframe}\eeq
This alternative frame-dragging equation now takes the same
form as the single-fluid equation derived by Hartle~\cite{H1}
(the difference being in the factor multiplying $\tilde{L}_{\p}$).
Thus we know that if we were to solve it numerically we would 
find that $\tilde{L}_{\p}$ decreases monotonically 
from the centre of the star to the surface. 
In other words, the results would be similar to those shown in 
figure~\ref{frame1}. In view of this we do not show such results
here.

\section{Conclusions}

In this paper we have developed a framework for constructing
and analyzing slowly rotating relativistic superfluid neutron stars. 
The model is based on  the standard two-fluid representation
for superfluids wherein all the charged components: protons, electrons
and lattice nuclei, are considered as a single fluid coexisting with 
the superfluid neutrons.
We have applied the formalism to a simple equation of state for which 
the two fluids are described by polytropes. We have then 
studied the effects of rotation on the resultant stars, 
with particular focus on the effects due to the fact that the 
two fluids need not rotate together. 

The present results provide a framework that opens the door
to fully relativistic studies of many important astrophysical
problems. 
We are currently extending this work in two directions. First, 
we are investigating the effects of entrainment on rotating stellar
models. Secondly, we are extending previous studies of 
quasinormal mode oscillations in relativistic superfluid stars
\cite{CLL}
to incorporate the effects of slow rotation. Of particular 
interest would be a study of the low-frequency inertial modes, 
following in the footsteps of Lockitch et al. \cite{laf00}.
The calculation of such modes has been brought into 
sharp focus since the discovery that the r-modes
are generically unstable due to the emission of gravitational 
waves. So far the only study of the r-modes in the superfluid 
context is the Newtonian work of Lindblom and Mendell \cite{LM00}. 
Our plan is to study the same problem within 
the framework of relativity. This is an issue that
demands immediate attention given the possibility
that the r-modes may lead to detectable gravitational 
waves and the  fact that a new generation of 
large scale interferometric detectors  will come on-line soon.

\acknowledgments

We want to thank John Friedman, David Langlois, Andreas Reisenegger 
and Mal Ruderman for useful 
discussions.  NA gratefully acknowledges  support by 
PPARC grant PPA/G/1998/00606 in the UK, as well as the great hospitality 
of the ITP in Santa Barbara 
where a part of this work was done (supported by NSF 
grant No. PHY94-07194).
GLC gratefully acknowledges partial 
support from a Saint Louis University SLU2000 Faculty Research Leave 
award, and the warm hospitality of l'Observatoire de Paris-Meudon, the 
University of Southampton and the Center for Gravitation and Cosmology 
of the University of Wisconsin at Milwaukee where various parts of this 
work were carried out.

\section{Appendix I}

Here we give the four combinations used in the right-hand-sides of the 
field equations that determine the four metric coeffcients
\begin{eqnarray}
    T^0_0 &=& \Lambda_{\rm o} - \mu_{\rm o} \n_{\rm o} \left(\eta - {r^2 
              {\rm sin}^2\theta \over e^{\nu}} \Omega_{\n} \tilde{L}_{\n}
              \right) - \chi_{\rm o} \p_{\rm o} \left(\Phi - {r^2 
              {\rm sin}^2\theta \over e^{\nu}} \Omega_{\p} \tilde{L}_{\p}
              \right) + \cr
           && \cr
           && {r^2 {\rm sin}^2\theta \over 2 e^{\nu}} \A_{\rm o} 
              \n_{\rm o} \p_{\rm o} \left(\Omega_{\n} - \Omega_{\p}
              \right)^2\ , \cr
           && \cr
    T^1_1 &=& \Psi_{\rm o} + \n_{\rm o} \left(\n_{\rm o} \left.\b00
              \right|_{\rm o} + \p_{\rm o} \left.\a00\right|_{\rm o}\right) 
              \eta + \p_{\rm o} \left(\p_{\rm o} \left.\c00\right|_{\rm o} 
              + \n_{\rm o} \left.\a00\right|_{\rm o}\right) \Phi + \cr 
           && \cr
           && {r^2 {\rm sin}^2\theta \over 2 e^{\nu}} \n_{\rm o} \p_{\rm o} 
              \left(\A_{\rm o} + \n_{\rm o} \left.{\partial \A \over \partial 
              \n} \right|_{\rm o} + \p_{\rm o} \left.{\partial \A \over 
              \partial \p}\right|_{\rm o} + 2 \n_{\rm o} \p_{\rm o} 
              \left.{\partial \A \over \partial x^2}\right|_{\rm o}\right) 
              \left(\Omega_{\n} - \Omega_{\p}\right)^2 \ , \cr
          && \cr
    T_{0 3} - {1 \over 2} T g_{0 3} &=& {1 \over 2} r^2 {\rm sin}^2\theta 
            \left(3 \Psi_{\rm o} - \Lambda_{\rm o}\right) \left(
            \tilde{L}_{\n} + \Omega_{\n}\right) - r^2 
            {\rm sin}^2\theta \left(\mu_{\rm o} \n_{\rm o} \Omega_{\n} + 
            \chi_{\rm o} \p_{\rm o} \Omega_{\p}\right) \ , \cr
          && \cr
    T^2_2 - T^3_3 &=& {r^2 {\rm sin}^2\theta \over e^{\nu}} \left(
            \n_{\rm o} \left(\B_{\rm o} \n_{\rm o} \Omega_{\n} + \A_{\rm o} 
            \p_{\rm o} \Omega_{\p}\right) \tilde{L}_{\n} + \p_{\rm o} 
            \left(\C_{\rm o} \p_{\rm o} \Omega_{\p} + \A_{\rm o} \n_{\rm o} 
            \Omega_{\n}\right) \tilde{L}_{\p}\right) \ . 
\end{eqnarray}
Some relationships useful for obtaining the stress-energy combinations just 
given are the following:
\beq 
    x^2 = \n_{\rm o} \p_{\rm o} \left(1 + \eta + \phi + { 1 \over 2} 
          \left(\omega_{\n} - \omega_{\p}\right)^2\right) \ ,
\eeq
\begin{eqnarray}
  \A &=& \A_{\rm o} + \n_{\rm o} \left(\left.{\partial \A \over \partial \n}
         \right|_{\rm o} + \p_{\rm o} \left.{\partial \A \over \partial x^2}
         \right|_{\rm o}\right) \eta + \p_{\rm o} \left(\left.{\partial \A 
         \over \partial \p}\right|_{\rm o} + \n_{\rm o} \left.{\partial \A 
         \over \partial x^2}\right|_{\rm o}\right) \Phi + \cr
      && \cr
      && {1 \over 2} \n_{\rm o} \p_{\rm o} \left.{\partial \A \over 
         \partial x^2}\right|_{\rm o} \left(\omega_{\n} - \omega_{\p}
         \right)^2 \ ,
\end{eqnarray}
with similar results for $\B$ and $\C$,
\begin{eqnarray}
  \Lambda &=& \Lambda_{\rm o} - \mu_{\rm o} \n_{\rm o} \eta - \chi_{\rm o} 
              \p_{\rm o} \Phi - {1 \over 2} \A_{\rm o} \n_{\rm o} 
              \p_{\rm o} \left(\omega_{\n} - \omega_{\p}\right)^2 \ , 
              \label{lam_pert}
\end{eqnarray}
and
\begin{eqnarray}
  \Psi &=& \Psi_{\rm o} + \n_{\rm o} \left(\n_{\rm o} \left.\b00
           \right|_{\rm o} + \p_{\rm o} \left.\a00\right|_{\rm o}\right) 
           \eta + \p_{\rm o} \left(\p_{\rm o} \left.\c00\right|_{\rm o} + 
           \n_{\rm o} \left.\a00\right|_{\rm o}\right) \Phi + \cr 
        && \cr
        && {1 \over 2} \n_{\rm o} \p_{\rm o} \left(\A_{\rm o} + \n_{\rm o} 
          \left.{\partial \A \over \partial \n}\right|_{\rm o} + \p_{\rm o} 
          \left.{\partial \A \over \partial \p}\right|_{\rm o} + 2
          \n_{\rm o} \p_{\rm o} \left.{\partial \A \over \partial x^2}
          \right|_{\rm o}\right) \left(\omega_{\n} - \omega_{\p}\right)^2
\end{eqnarray}
where $\Psi_{\rm o} - \Lambda_{\rm o} = \mu_{\rm o} \n_{\rm o} + 
\chi_{\rm o} \p_{\rm o}$.  The analogous expressions for the Einstein and 
Ricci tensor components can be found in Hartle \cite{H1}.


\begin{references}
\bibitem{newcoor}Strictly speaking, this new coordinate is not orthogonal 
in the sense that it re-introduces the $g_{1 2}$ metric coefficient.  
However, this coefficient is completely decoupled from the other metric 
and matter variables to the order we are working, and so can be ignored 
(at least for the applications we have considered).
\end{references}
\end{document}